\documentclass[aps,preprintnumbers,amsmath,amssymb,latexsym,array,twocolumn,superscriptaddress,nofootinbib]{revtex4-1}

\usepackage{amssymb}
\usepackage{amsmath}
\usepackage{epsfig}
\usepackage{hyperref}
\usepackage{breakurl}
\usepackage{xcolor}

\usepackage{extarrows}
\usepackage{shuffle}

\usepackage{graphicx}
\usepackage{subfig}

\usepackage{overpic}

\newcommand{\LL}{{\scriptscriptstyle{\rm LL}}}

\graphicspath{{./artbook/}}

\begin{document}

\title{Analytic Four-Point Lightlike Form Factors and OPE of Null-Wrapped Polygons}

\author{Yuanhong Guo}
\email{guoyuanhong@itp.ac.cn}
\affiliation{CAS Key Laboratory of Theoretical Physics, Institute of Theoretical Physics, Chinese Academy of Sciences,  Beijing, 100190, China}
\affiliation{School of Physical Sciences, University of Chinese Academy of Sciences, Beijing 100049, China}
\author{Lei Wang}
\email{wanglei1@bhu.edu.cn}
\affiliation{College of Physical Science and Technology, Bohai University, Jinzhou 121013, China}
\affiliation{CAS Key Laboratory of Theoretical Physics, Institute of Theoretical Physics, Chinese Academy of Sciences,  Beijing, 100190, China}
\affiliation{School of Physical Sciences, University of Chinese Academy of Sciences, Beijing 100049, China}
\affiliation{School of Physics, Peking University, Beijing 100871, China }
\affiliation{Center for High Energy Physics, Peking University, Beijing 100871, China}
\author{Gang Yang\vspace{2mm}}
\email{yangg@itp.ac.cn}
\affiliation{CAS Key Laboratory of Theoretical Physics, Institute of Theoretical Physics, Chinese Academy of Sciences,  Beijing, 100190, China}
\affiliation{School of Physical Sciences, University of Chinese Academy of Sciences, Beijing 100049, China}
\affiliation{School of Fundamental Physics and Mathematical Sciences, Hangzhou Institute for Advanced Study, UCAS, Hangzhou 310024, China}
\affiliation{International Centre for Theoretical Physics Asia-Pacific, Beijing/Hangzhou, China}
\affiliation{Peng Huanwu Center for Fundamental Theory, Xian 710127, China}


\begin{abstract}

We obtain for the first time the analytic two-loop four-point MHV lightlike form factor of the stress-tensor supermultiplet in planar ${\cal N}=4$ SYM where the momentum $q$ carried by the operator is taken to be massless. Remarkably, we find that the two-loop result can be constrained uniquely by the infrared divergences and the collinear limits using the master-bootstrap method.
Moreover, the remainder function depends only on three dual conformal invariant variables,
which can be understood from a hidden dual conformal symmetry of the form factor arising in the lightlike limit of $q$.
The symbol alphabet of the remainder contains only nine letters, which are closed under the action of the dihedral group $D_4$.
Based on the dual description in terms of periodic Wilson lines (null-wrapped polygons), we also consider a new OPE picture for the lightlike form factors and introduce a new form factor transition that corresponds to the three-point lightlike form factor.
With the form factor results up to two loops,
we make some all-loop predictions using the OPE picture.

\end{abstract}

\maketitle

\section{Introduction}

\noindent
Significant progress has been made in past ten years about scattering amplitudes and correlation functions in gauge theories. In particular, in  ${\cal N}=4$ supersymmetric Yang-Mills (SYM) theory, both high-loop and non-perturbative results have been obtained and various dualities have also been revealed, see \emph{e.g.}~\cite{Alday:2008yw, Arkani-Hamed:2010zjl, Beisert:2010jr, Dixon:2013uaa, Elvang:2013cua,  Henn:2014yza, Travaglini:2022uwo} for review.
The study of form factors (FFs), the key quantities representing the overlap between on-shell states and local operators, has provided further deep insights into the symmetries and dynamics of gauge theories.

The FF of the stress-tensor supermultiplet in ${\cal N}=4$ SYM theory
has been a particularly rich playground for revealing hidden simplicities and structures, see \cite{Yang:2019vag} for a recent review.
The duality between this FF and the periodic Wilson loop (WL) was first proposed at strong coupling \cite{Alday:2007he,Maldacena:2010kp} and then checked at one-loop order \cite{Brandhuber:2010ad} (see also \cite{Brandhuber:2011tv, Ben-Israel:2018ckc, Bianchi:2018rrj}).
Recently, the FF operator product expansion (OPE) was developed \cite{Sever:2020jjx, Sever:2021nsq, Sever:2021xga},
extending the OPE for amplitudes \cite{Alday:2010ku, Basso:2013vsa, Basso:2013aha,Basso:2014koa,Basso:2014nra,Basso:2014hfa,Basso:2015rta,Basso:2015uxa,Belitsky:2014sla,Belitsky:2014lta,Belitsky:2016vyq}.
With the FFOPE prediction, the three-point FF has been computed to the remarkable eight loops \cite{Dixon:2020bbt, Dixon:2022rse}, using the symbol bootstrap method \cite{Dixon:2011pw, Brandhuber:2012vm, Dixon:2013eka,Dixon:2014iba,Golden:2014pua,Drummond:2014ffa, Caron-Huot:2016owq,Dixon:2016nkn, Drummond:2018caf, Caron-Huot:2019vjl,Dixon:2020cnr, Zhang:2019vnm, He:2020vob, Golden:2021ggj}.
Surprisingly, these new results reveal an intriguing duality between the three-point FF and six-point amplitude \cite{Dixon:2021tdw}; a connection between the two quantities was first noted at two loops in \cite{Brandhuber:2012vm}.
While these studies mostly focus on the three-point FF, it is important to explore more general FFs.

In this paper we initiate the study of \emph{lightlike form factors} of the stress-tensor supermultiplet in which the momentum of the local operator $q$ is taken to be lightlike.
Due to the special lightlike condition, this FF presents intrinsically new hidden structure and simplicity.
The lightlike FFs are expected to be dual to polygonal WLs with a lightlike period $x_{i+n}-x_i=q$, which will be also referred to as \emph{null-wrapped polygons}.
Unlike the case with off-shell $q$, the duality for the lightlike FFs, if true, should imply an exact directional dual conformal symmetry (DDCS) of the FFs along the $q$ direction.
This symmetry was checked at the integrand level for the three-point FF of the stress-tensor multiplet \cite{Lin:2021lqo}.
For the integrated FFs, the DDCS suggests that the finite remainder function of an $n$-point lightlike FF depends on $3n-9$ independent dual conformal invariant ratios (similar to the counting for amplitudes \cite{Drummond:2007au}).
Thus the first non-trivial case is the four-point case which depends on three independent ratios.

We will present a first analytic computation of the two-loop four-point lightlike FF
using the master-integral (MI) bootstrap method \cite{Guo:2021bym,Guo:2022pdw}.
Remarkably, we find the constraints of infrared (IR) divergences and collinear limits
can uniquely determine the two-loop finite remainder of the FF.
Moreover, we confirm that the finite remainder depends only on three independent ratios. This shows explicitly the DDCS of the FF 
and provides the most compelling verification for the conjectured duality between the FF and periodic WL. 

In terms of the dual null-wrapped polygon, we propose a non-perturbative lightlike FFOPE framework.
Due to the special lightlike condition, our formalism contains new properties
comparing to previous studies \cite{Basso:2013vsa,Sever:2020jjx}.
A novel lightlike FF transition is involved in our picture, which originates from the three-point lightlike FF.
The regularized finite ratio function is introduced and the explicit parametrization of the FFOPE parameters is provided.
Combining the FFOPE picture and the perturbative results, we are able to make some all-order predictions.
Our method also provides a new integrability framework to explore the WLs with lightlike periods.

\section{Two-loop computation via bootstrap}
\label{sec:ansatz}

\noindent
The FF we consider is defined as
\begin{equation}
{\cal F}^{\LL}_{4}
= \int d^D x \mathrm{e}^{-\mathrm{i} q\cdot x} \langle p_1, p_2, p_3, p_4 | {\cal L}(x) | 0 \rangle \big|_{q^2=0},
\end{equation}
where $p_i (p_i^2=0)$ are on-shell momenta, $q = \sum_{i=1}^4 p_i$, and the superscript ``LL" indicates the lightlike condition of $q$.
Without loss of generality, we choose the operator as the chiral Lagrangian ${\cal L} = {\rm tr}(F_{\alpha\beta} F^{\alpha\beta}) + \cdots$ in the stress-tensor multiplet.

Before constructing the two-loop ansatz, we first consider the tree-level and one-loop results.
The tree-level MHV FF takes the simple form \cite{Brandhuber:2011tv}
\begin{align}
	{\cal F}^{\LL, (0)}_4 = \frac{\delta^{(8)}(\sum_i \lambda_i \eta_i)}{\langle12\rangle \langle23\rangle \langle34\rangle \langle41\rangle} \,.
\end{align}
We apply the $D=4-2\epsilon$ dimensional unitarity-cut method \cite{Bern:1994zx, Bern:1994cg, Britto:2004nc} to obtain the one-loop result that is valid to all orders in $\epsilon$ expansion \footnote{Up to ${\cal O}(\epsilon^0)$ order it matches the result in \cite{Brandhuber:2010ad}. Here we also need to know higher ${\cal O}(\epsilon)$ result since it is necessary for subtracting the IR divergence of higher loop by the BDS expansion we will mention later.}:
\begin{align}
	\label{eq:F1loopinGexp}
	{\cal F}^{\LL, (1)}_4 = {\cal F}^{\LL, (0)}_4 \hat{{\cal F}}_4^{\LL, (1)} = {\cal F}^{\LL, (0)}_4 \Big( {\cal G}_1^{(1)} +  B \, {\cal G}_2^{(1)} \Big)\,,
\end{align}
where $B$ is a parity-odd factor
\begin{align}\label{eq:defB}
	B \equiv \frac{s_{12}s_{34} +s_{23}s_{14} -s_{13}s_{24}}{4 \mathrm{i} \varepsilon(1234)} \,,
\end{align}
with $\varepsilon(1234) \equiv \varepsilon_{\mu \nu \rho \sigma} p_1^{\mu} p_2^{\nu} p_3^{\rho} p_4^{\sigma}$, and ${\cal G}^{(1)}_a$ can be given in terms of the bubble, box and pentagon MIs; see Appendix~\ref{app:oneloop} for details.

\begin{figure}[tb]
  \centering
  \includegraphics[width=1.\linewidth]{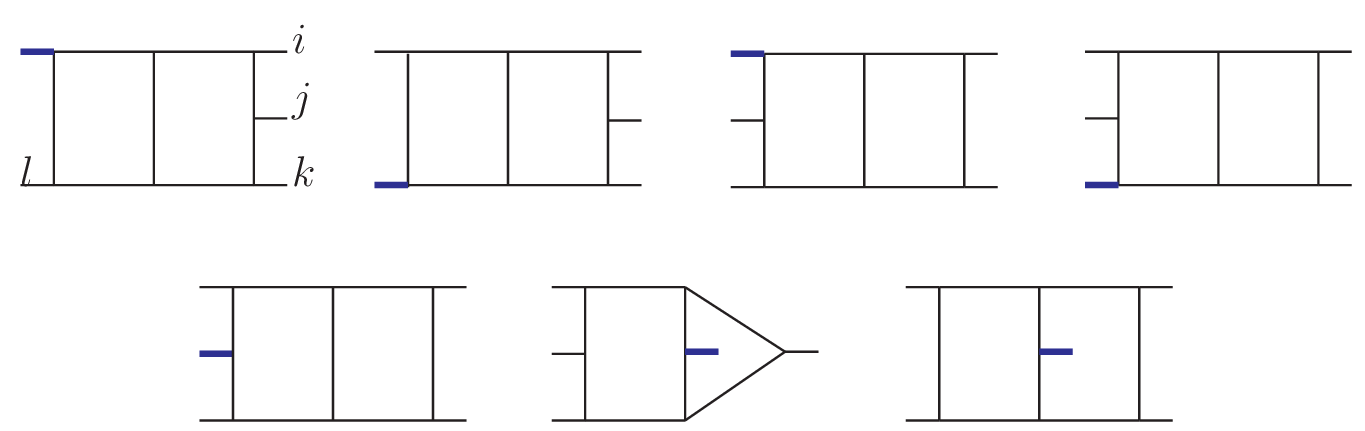}
  \caption{Topologies of maximal number of propagators for the FF, where the blue leg carries momentum $q$ and external leg configurations are cycling $\{p_1, p_2, p_3, p_4\}$.}
  \label{fig:twolooptopology}
\end{figure}

A natural guess is that the two-loop result takes a similar structure as one loop \footnote{Our choice of the ansatz is based on the following reasons. First, the two-loop FF must contain the spinor factor $B$ in order to satisfy the infrared structure given by BDS ansatz. Second, if there are extra spinor structures beyond $B$, it has to be at finite order. Furthermore, the WL should contain purely even functions, thus the duality between the FF and WL indicates that the finite remainder should be pure functions without even $B$ factor (which is indeed true as we discuss later). In any case, we stress that the correctness of our ansatz can be always verified by simple unitarity checks which we have checked and will be mentioned below.}, we propose the ansatz of two-loop planar FF as
\begin{equation}
	\label{eq:F2loopAnsatz}
	{\cal F}^{\LL, (2)}_4 = {\cal F}^{\LL, (0)}_4 \hat{{\cal F}}_4^{\LL, (2)} = {\cal F}^{\LL, (0)}_4 \Big( {\cal G}_1^{(2)} +  B \, {\cal G}_2^{(2)} \Big)\,,
\end{equation}
where ${\cal G}_a^{(2)}$ are pure uniformly transcendental (UT) functions expanded in terms of a set of the two-loop MIs.
The topologies with maximal numbers of propagators are shown in Figure~\ref{fig:twolooptopology}.
Then the most general ansatz contains $590$ MIs for each ${\cal G}_a^{(2)}$:
\begin{equation}
    \label{eq:ansatz2loop}
    {\cal G}_a^{(2)} = \sum_{i=1}^{590} c_{a,i} I_{i}^{(2)} \,,
\end{equation}
where $c_{a, i}$ are the coefficients to be solved.
If we choose the MIs as the pure UT integrals, which have been constructed in \cite{Papadopoulos:2015jft, Gehrmann:2015bfy, Chicherin:2017dob, Abreu:2018aqd, Gehrmann:2018yef, Abreu:2018rcw, Chicherin:2018mue,Chicherin:2018old, Chicherin:2020oor} using the differential equations method \cite{Henn:2013pwa}, the coefficients $c_{a,i}$ should be rational numbers. Note that we have separated the $B$ factor from $c_{a,i}$ in our ansatz.

Next we apply a series of physical constraints to determine the ansatz coefficients $c_{a,i}$, including the dihedral group $D_4$ (the symmetry of the FF under cyclicly permuting and flipping external momenta), the IR subtraction and the collinear limit, as in~\cite{Guo:2021bym,Guo:2022pdw}.

The two major constraints are IR divergences \cite{Catani:1998bh, Sterman:2002qn} and collinear factorization \cite{Bern:1993qk, Bern:1994zx, Kosower:1999xi}.
For the planar amplitudes or FFs in ${\cal N}=4$ SYM, a convenient representation to capture both the IR and collinear behavior
is the BDS expansion \cite{Bern:2005iz, Anastasiou:2003kj}, which at two-loop gives:
\begin{equation}
    \label{eq:BDSansatz}
    \hat{{\cal F}}_4^{(2)} = {1\over2} \big( \hat{{\cal F}}_4^{(1)}(\epsilon) \big)^2 + f^{(2)}(\epsilon) \hat{{\cal F}}_4^{(1)}(2\epsilon) + {\cal R}_4^{(2)} + {\cal O}(\epsilon) \,,
\end{equation}
where $f^{(2)}(\epsilon)=-2 \zeta_{2}-2 \zeta_{3} \epsilon-2 \zeta_{4} \epsilon^{2}$.
The finite remainder function ${\cal R}$ has the nice collinear behavior
\begin{equation}
    \label{eq:4ptremainderCL}
    {\cal R}_{\textrm{4}}^{\LL, (2)} \ \xlongrightarrow{\mbox{$p_i \parallel p_{i+1}$}} \ {\cal R}_{\textrm{3}}^{\LL, (2)} = -6 \zeta_4 \,,
\end{equation}
where as mentioned in the introduction, the remainder of the three-point lightlike FF is a transcendental number.

To simplify the calculation, we can consider the ``symbol" of MIs, which was introduced to the study of the two-loop six-gluon amplitude \cite{Goncharov:2010jf} and will significantly simplify the transcendental functions by mapping them into tensor products of function arguments, for simple examples:
${\cal S}(\log (x)) = x, {\cal S}({\rm Li}_2(x)) = -(1-x) \otimes  x$.
Substituting the MI symbol results into our ansatz \eqref{eq:F2loopAnsatz}, we obtain an $\epsilon$-expansion form of the FF \footnote{Since the FF is uniformly transcendental, the tensor degree at a given order in $\epsilon$-expansion is fixed, \emph{e.g.}~the finite order has degree $k=4$.}:
\begin{equation}
    \label{eq:symF4}
    {\cal S}(\hat{{\cal F}}_4^{\LL, (2)}) = \sum_{k\geq0} \epsilon^{k-4} \sum_I \alpha_I(c) \otimes_{i=1}^k w_{I_i} \,,
\end{equation}
where $w_I$ are rational functions of kinematic variables and are called symbol \emph{letters},
and $\alpha_I(c)$ are linear combinations of $c_{a,i}$ in \eqref{eq:ansatz2loop} (up to the factor $B$).
There are 31 independent letters for the MIs we consider and we review them in Appendix~\ref{app:symbolletter}. 
We point out that the symbol may miss some terms containing transcendental numbers. Such terms can be recovered by considering further the FF at function-level; see Appendix~\ref{app:numericalDDCI} for further discussion and checks. 

We summarize the parameters after each constraint in Table~\ref{tab:masterbootstrap}.
Remarkably, the remaining $21$ parameters after collinear constraint all contribute to ${\cal O}(\epsilon)$ order.
These terms can be safely ignored if one is only interested in up to the finite order of the two-loop FF.
In other words, the two-loop finite remainder can be uniquely fixed by the IR and collinear constraints.
To complete and cross-check the result, we also apply a spanning set of $D$-dimensional unitarity cuts \cite{Bern:1994zx, Bern:1994cg, Britto:2004nc} which verify the bootstrap result.

We provide the two-loop FF expanded in terms of MIs \eqref{eq:ansatz2loop} in the ancillary files.

\begin{table}[t]
\centering
\vskip .1 cm
\begin{tabular}{| l | c |}
\hline
Constraints  			        &  Parameters left    \\ \hline \hline
Starting ansatz                 &  $590 \times 2$ \\ \hline
Symmetries of external legs 	&  $168$  \\ \hline
IR (Symbol)   			        &  $109$  \\ \hline
Collinear limit (Symbol)    	&  $43$  \\ \hline
IR (Function)			    	&  $39$ \\ \hline
Collinear limit (Function)		&  $21$ \\ \hline
Keeping up to $\epsilon^0$ order (or via unitarity) 	&  $0$ \\ \hline
\end{tabular}
\caption{Solving for parameters via master bootstrap.
\label{tab:masterbootstrap}
}
\end{table}

\section{Two-loop finite remainder}
\label{sec:remainder}

\noindent
The two-loop finite remainder ${\cal R}_{\textrm{4}}^{\LL, (2)}$ presents several nice properties.
First, all the terms which are proportional to the $B$ factor \eqref{eq:defB} cancel in the two-loop finite remainder function, since
\begin{align}
    {\cal G}_2^{(2)}(\epsilon) = {\cal G}_1^{(1)}(\epsilon) \, {\cal G}_2^{(1)}(\epsilon)+f^{(2)}(\epsilon) \, {\cal G}_2^{(1)}(2\epsilon) + {\cal O}(\epsilon) \,.
\end{align}
The remaining parity-odd part from ${\cal G}_1^{(2)}(\epsilon)$ also cancels, thus the full remainder is purely even.
These cancellations are consistent with the conjectured duality between the FF and periodic WL, since the latter is purely even.
Similar properties was also noted for the six-gluon amplitude \cite{Bern:2008ap}.

\begin{figure}[t]
  \centering
  \includegraphics[scale=0.37]{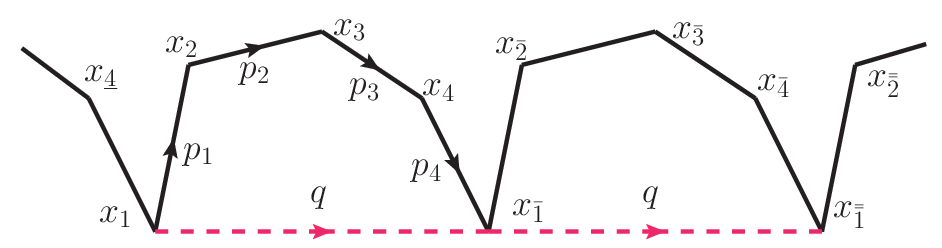}
  \caption{Dual periodic Wilson line.}
  \label{fig:WL4pt}
\end{figure}

Second, the remainder function only depends on three ratio variables,
\begin{align}\label{eq:3DDCIratios}
	& u_1 \equiv \frac{s_{12}}{s_{34}} = \frac{x_{13}^2}{x_{3\bar{1}}^2} \,, \quad
	u_2 \equiv \frac{s_{23}}{s_{14}} = \frac{x_{24}^2}{x_{4\bar{2}}^2} \,, \\
	& u_3 \equiv \frac{s_{123} s_{134}}{s_{234} s_{124}} = \frac{x_{14}^2 x_{3\bar{2}}^2}{x_{2\bar{1}}^2 x_{4\bar{3}}^2} \,, \nonumber
\end{align}
where $x$ are defined in the dual momentum space \cite{Drummond:2006rz} as shown in Figure~\ref{fig:WL4pt}, and satisfy $x_{i+1} - x_i = p_i $ and $x_{ij} = x_j - x_i$.
Notably, these variables satisfy a special dual conformal symmetry in the dual $x$ space \cite{Lin:2021lqo, Bern:2018oao, Chicherin:2018wes}:
\begin{align}
    \delta_{q} x_{i}^{\mu} \equiv \frac{1}{2} x_{i}^{2} q^{\mu}-\left(x_{i} \cdot q\right) x_{i}^{\mu} \,,
\end{align}
which is the DDCS mentioned in the introduction.
Our result shows for the first time that the lightlike FF has the exact DDCS at the {integrated}-level.
Further discussion about the DDCS is given in Appendix~\ref{app:numericalDDCI}.
We stress again that the emergence of DDCS is highly nontrivial, as we did not impose any DDCS in our bootstrap computation \footnote{If we do use DDCI as an input constraint and apply it at  the symbol-level in Table~\ref{tab:masterbootstrap}, we find that it can reduce the $10$ of $43$ parameters after the collinear constraint.
In the symbol bootstrap, the DDCS can be used at the beginning to constrain the alphabet in the symbol. See further discussion in the Outlook.}.

Third, the DDCS also leads to the compact two-loop remainder. We illustrate this at symbol-level.
The remainder has uniform transcendentality-degree 4 and its symbol is
\begin{equation}
	\label{eq:remaindersymbol}
	{\cal S}({\cal R}_4^{\LL, (2)}) = \sum_{i=1}^{1283} {\tilde c}_i \, w_{i_1} \otimes w_{i_2} \otimes w_{i_3} \otimes w_{i_4} \,,
\end{equation}
where ${\tilde c}_i$ are rational numbers.
Nicely, the set of letters $\{w_i\}$ has only 9 elements:
\begin{equation}\label{eq:alphabet}
    w_i \in \{u_a \,, \ 1-u_a \,, \ x_{1234} \,, \ x_{234q} \,, \ x_{123q}\},
\end{equation}
where $a=1,2,3$ and
\begin{equation}\label{eq:3Xratios}
	x_{ijkl} = \frac{s_{ij} s_{kl} +s_{il} s_{jk} -s_{ik} s_{jl} +4\mathrm{i}\varepsilon(ijkl)}{s_{ij} s_{kl} +s_{il} s_{jk} -s_{ik} s_{jl} -4\mathrm{i}\varepsilon(ijkl)} .
\end{equation}
They are closed under the action of the dihedral group $D_4$.
Note that the $x$ are parity-odd variables and can be expressed as functions of $u_a$ in \eqref{eq:3DDCIratios}; see Appendix~\ref{app:symbolletter} for details.
These are precisely the nice ones satisfying DDCS among the 31 letters in two-loop MIs.

We provide the two-loop remainder at both the symbol- and function-level in the ancillary files.

\section{Lightlike FFOPE}
\label{sec:OPE}

\noindent
We now propose a non-perturbative framework for the lightlike FFs using the dual null-wrapped WL picture.
While we will consider the OPE similar in the framework of \cite{Basso:2013vsa,Sever:2020jjx},
due to the special lightlike condition and the DDCS property, our construction requires a genuinely new formulation.
In particular, a new building block—the lightlike FF transition—appears, which is associated with the three-point lightlike FF. 
Below we focus on the four-point case and the generalization to higher points is straightforward.

\begin{figure}[t]
  \centering
  \includegraphics[scale=0.38]{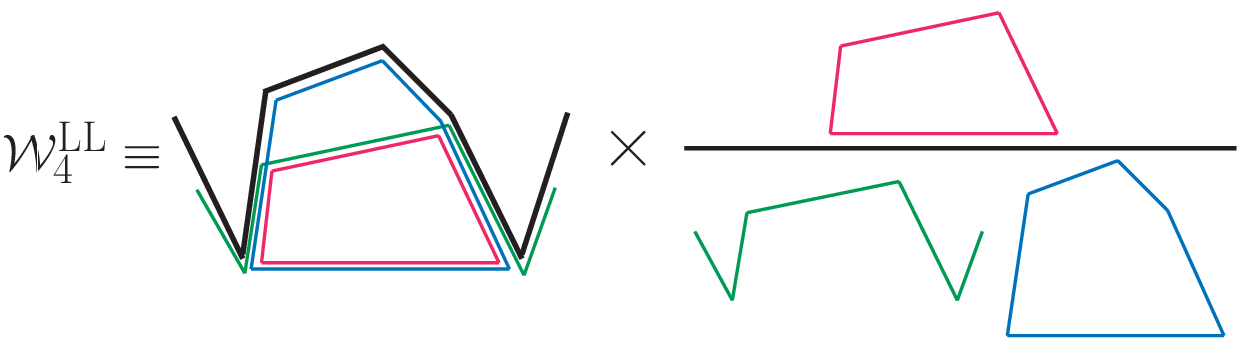}
  \caption{Regularization of the null-wrapped polygon.}
  \label{fig:OPEregularize}
\end{figure}

As shown in Figure~\ref{fig:WL4pt}, the $4$-point lightlike FF is dual to a four-sided periodic null-polygon.
The first important step is to define the regularized four-point null-wrapped polygon as the following finite ratio
\begin{equation}\label{eq:W4ratio}
	{\cal W}^\LL_{4} = {\hat {\cal F}^\LL_{4} \times {\cal W}_{\textrm{sq}} \over
	\hat{\cal F}^\LL_{3} \times {\cal W}_{\textrm{pen}} } \,,
\end{equation}
where $W_{\textrm{sq}}$ and $W_{\textrm{pen}}$ are the null square and pentagon WLs, and  $\hat{\cal F}^\LL_{n} = {\cal F}^\LL_{n}/{\cal F}^{\LL,(0)}_{n}$.
Importantly, the three-point lightlike FF plays the role of a vacuum.
This subtraction is illustrated in Figure~\ref{fig:OPEregularize}.

Expanding \eqref{eq:W4ratio} perturbatively, the one-loop finite ratio is obtained as
\begin{align}\label{eq:W1loopOPE}
{\cal W}^{\LL, (1)}_{4} & =-\frac{1}{2} \log \left(\frac{(1-u_1) (1-u_2)}{1-u_3}\right) \\
 & \qquad \times \log \left(\frac{(1-u_1) (1-u_2) u_3}{(1-u_3)u_1 }\right), \nonumber
\end{align}
which depends only on the three ratios in \eqref{eq:3DDCIratios} and provides a non-trivial check for the DDCI.

Beyond one loop, the finite ratio ${\cal W}^\LL_{n}$ is related to the BDS remainder ${\cal R}^\LL_n$ as
(see also \cite{Dixon:2020bbt, Sever:2020jjx})
\begin{equation}\label{eq:WtoR}
{\cal W}^\LL_{n} = \exp \bigg[ {\Gamma_{\text{cusp}} \over 4} {\cal W}^{\LL, (1)}_{n} + {\cal R}^\LL_n \bigg]  \,,
\end{equation}
where $\Gamma_{\rm cusp}= 4g^2+..$ is the cusp anomalous dimension \cite{Korchemsky:1985xj, Korchemsky:1988si, Beisert:2006ez}.
Using the previously computed ${\cal R}_4^{\LL, (2)}$, we obtain the two-loop result ${\cal W}^{\LL,(2)}_{4}$.
We comment here that one can also consider ``BDS-like remainders" by subtracting different one-loop expressions likes \eqref{eq:WtoR}, as in \cite{Dixon:2020bbt, Dixon:2022rse}.
We give more details in Appendix~\ref{app:BDSlike}.

Given the finite ratio \eqref{eq:W4ratio}, we propose the OPE for the lightlike FF as
\begin{equation}
	{\cal W}^\LL_{4} = \sum_{\psi} \mathrm{e}^{-E_\psi \tau +\mathrm{i}p_\psi \sigma +\mathrm{i}m_\psi \phi} \, {\mathbb P}(0 | \psi) \, {\mathbb F}^\LL(\psi) \,, \label{eq:OPELLF4}
\end{equation}
where the sum is over states $\psi$ carrying charges $\{E, p, m\}$, \emph{i.e.}~the GKP energy, momentum and angular momentum \cite{Alday:2010ku}. ${\mathbb P}(0 | \psi)$ is the pentagon transition \cite{Basso:2013vsa}, and ${\mathbb F}^\LL(\psi)$ is the new \emph{lightlike FF transition} which is associated to the three-point FF vacuum.

\begin{figure}[b]
	\centering
	\begin{overpic}[scale=0.39]{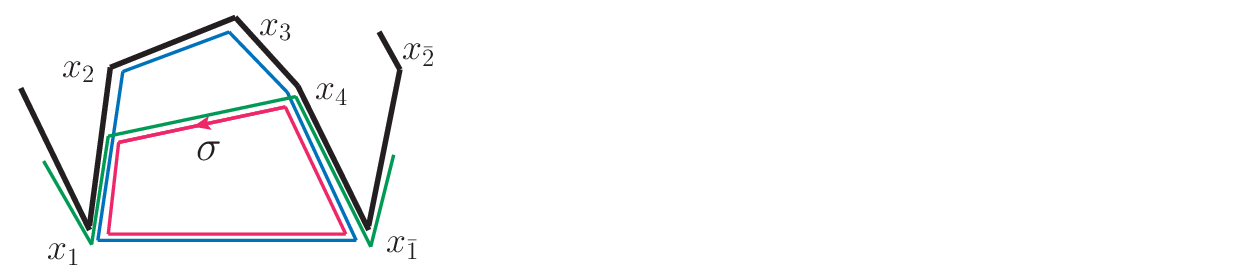}
		\put(37,17){
		$u_1 = \frac{x^2_{13}}{x^2_{3\bar{1}}} = \mathrm{e}^{-2\sigma}$,}
		\put(69,17){
		$u_2 = \frac{x^2_{24}}{x^2_{4\bar{2}}} = \mathrm{e}^{-2\tau}$}
		\put(37,5){
		$u_3 = \frac{x^2_{14} x^2_{3\bar{2}}}{x^2_{2\bar{1}} x^2_{4\bar{3}}} = \frac{ \cosh(\sigma-\tau) + \cos(\phi)}{\cosh(\sigma+\tau) + \cos(\phi)}$}
	\end{overpic}
	\caption{Parametrization of $\{\tau, \sigma, \phi\}$ for cross ratios.}
	\label{fig:OPEparametrization}
\end{figure}

The parameters $\{\tau, \sigma, \phi\}$ in \eqref{eq:OPELLF4} are related to the three ratios $u_i$ in \eqref{eq:3DDCIratios}. They parameterize the symmetries of the null square (in green color) and can be given in terms of dual coordinates as in Figure~\ref{fig:OPEparametrization}.
The OPE limit is obtained by taking $\tau\rightarrow \infty$ which corresponds to the limit of $p_2 \parallel p_3$.
Details of the parametrization are explained in Appendix~\ref{app:OPEparameter}.

\begin{figure}[t]
  \centering
  \includegraphics[scale=0.38]{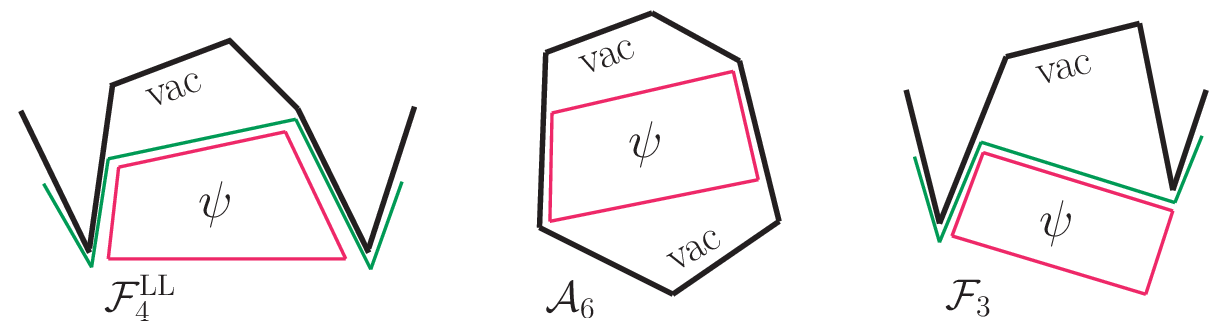}
  \caption{OPE decomposition for ${\cal F}_4^\LL$, ${\cal A}_6$, and ${\cal F}_3$.}
  \label{fig:OPEA6F3F4}
\end{figure}

We would like to comment on the novelty of our lightlike FFOPE \eqref{eq:OPELLF4}
by making an analogy with the known six-point amplitude \cite{Basso:2013vsa}
and three-point FF with $q^2\neq0$ \cite{Sever:2020jjx} (see Figure~\ref{fig:OPEA6F3F4}):
\begin{align}
	{\cal W}_{{\cal A}_{6}} & = \sum_{\psi} \mathrm{e}^{-E_\psi \tau +\mathrm{i} p_\psi \sigma +\mathrm{i} m_\psi \phi} \, {\mathbb P}(0 | \psi) \, {\mathbb P}(\psi|0) \,, \nonumber \\
	{\cal W}_{{\cal F}_{3}} & = \sum_{\psi} \mathrm{e}^{-E_\psi \tau +\mathrm{i} p_\psi \sigma} \, {\mathbb P}(0 | \psi) \, {\mathbb F}(\psi) \,.  \label{eq:OPEA6F3F4}
\end{align}
It should be clear that ${\mathbb F}^\LL(\psi)$ in \eqref{eq:OPELLF4} represents a novel building block.
Furthermore, the lightlike FFOPE distinguishes itself from that of \cite{Sever:2020jjx} by following reasons.
At the kinematics level, the lightlike limit  would reduce only one degree of freedom, but as our results explicitly show, in the limit the number of ratios is reduced by two.
The reason behind this is the emergence of a non-trivial hidden dual conformal symmetry.
This suggests that one should apply the lightlike condition from the very beginning to formulate the OPE.
In other words, the WL with lightlike periodic should be taken as a genuinely new configuration,
and our OPE proposal involving the novel transition ${\mathbb F}^\LL$ represents an independent formalism in parallel with previous ones. In Appendix~\ref{app:comparedifferentOPE}, we compare our lightlike FFOPE with other formalisms in more detials.

%

Below we utilize \eqref{eq:OPELLF4} and the perturbative results, which will lead to non-trivial all-loop predictions.
The large-$\tau$ expansion of ${\cal W}^\LL_{4}$ at one and two loops is straightforward to obtain:
\begin{align}
	\label{eq:W4OPEexpansion1}
	{\cal W}^{\LL, (1)}_{4} & = -\mathrm{e}^{-2 \tau} [1+ \cos(2\phi)] +{\cal O}(\mathrm{e}^{-3\tau}) \,, \\
	{\cal W}^{\LL, (2)}_{4} & = 2 \mathrm{e}^{-\tau} \cos(\phi) h_0^{(2)}(\sigma) + {\cal O}(\mathrm{e}^{-2\tau}) \,, 	\label{eq:W4OPEexpansion2}
\end{align}
where
\begin{align}
	h_{0}^{(2)}(\sigma) = & 4 \mathrm{e}^\sigma\big[ -{\rm Li}_3(\mathrm{e}^{-2\sigma}) + {\rm Li}_2(1-\mathrm{e}^{-2\sigma})  \\
	& \qquad -\sigma {\rm Li}_2(\mathrm{e}^{-2\sigma}) -\zeta_2 \big] +(\sigma \rightarrow -\sigma) \,. \nonumber
\end{align}

Similar to the six-gluon amplitude \cite{Alday:2010ku},
the leading contribution of order $\mathrm{e}^{-\tau}$ should come from the lightest state: the gluonic excitation states $F \equiv F_{z^-}$ and $\bar{F} \equiv F_{\bar{z}^-}$.
This is also different from the FFOPE in \cite{Sever:2020jjx} where the first excitation is the double-scalar state.
We find that the one-loop result starts at $\mathrm{e}^{-2\tau}$ while two-loop starts at $\mathrm{e}^{-\tau}$.
This suggests that the transition for the single gluonic excitation state contributes from two-loop order. By matching \eqref{eq:W4OPEexpansion2} with \eqref{eq:OPELLF4}, one obtains the leading order lightlike FF transition as
\begin{equation}
	\label{eq:FFtransition}
	{\mathbb F}_{F/\bar{F}}^{\LL}(u)\big|_{g^2} = \frac{2\pi}{(u^2 +\frac{1}{4}) \cosh(\pi u)} \,, 
\end{equation}
where $u$ is the Bethe rapidity charactering the excitation state $\psi$~\cite{Basso:2013vsa}.

The new result \eqref{eq:FFtransition}, together with other known functions in \eqref{eq:OPELLF4}, allows us to make following predictions at any $\ell$-loop order:
\begin{align}
	& {\cal W}^{\LL, (\ell)}_{4} |_{ \tau^{\ell-1} \mathrm{e}^{-\tau}} = 0 \,, 	\label{eq:OPEprediction1} \\
	& {\cal W}^{\LL, (\ell)}_{4} |_{\tau^{\ell-2} \mathrm{e}^{- \tau }} = 2\cos(\phi) \int_{\mathbb{R}} \frac{du}{2\pi} \mathrm{e}^{2iu\sigma} \nonumber \\
	& \quad \times  \frac{\big( -E_{3/2}^{(1)}(u) \big)^{\ell-2}}{(\ell-2)!}
	\bigg[\frac{-2\pi^2}{(u^2 +\frac{1}{4})^2 \cosh^2(\pi u)}\bigg]
	\,, 	\label{eq:OPEprediction2}
\end{align}
where $E_s^{(1)}(u) = 2 \big( \psi(s +iu) +\psi(s -iu) -2\psi(1) \big)$
is the one-loop GKP energy~\cite{Basso:2010in},
and $\psi(x) = \partial_x \log \Gamma(x)$ is Euler's digamma function.
We provide more details and some generalization in Appendix~\ref{app:OPEdetail}.

\section{Discussion}
\label{sec:outlook}

\noindent
In this paper we obtain the two-loop four-point lightlike FF of the stress-tensor multiplet in ${\cal N}=4$ SYM which presents a remarkably simple form. We show explicitly that the lightlike FF possesses a hidden dual conformal symmetry, providing the strongest check of the duality between the FF and periodic WL to date.
We further develop the lightlike FFOPE framework where the lightlike condition induces some genuinely new features compared to previous formalism.
A key new element is the lightlike FF transition.
Combining our perturbative results and FFOPE, we make predictions at all-loop order.
The results presented in this work thus establish a foundation for non-perturbative studies of lightlike FFs and dual null-wrapped WLs.

Our study leaves multiple future directions to explore and we mention a few below.

I) One important problem is to determine the lightlike FF transition ${\cal F}^\LL(\psi)$ non-perturbatively via integrability \cite{Basso:2013vsa, Sever:2020jjx}.
The OPE date will help determine the high-loop four-point FF remainder via symbol bootstrap as in the three-point case \cite{Dixon:2020bbt, Dixon:2022rse}.
A preliminary study for the symbol bootstrap up to three loops is presented in Appendix~\ref{app:symbolbootstrap}.

II) Inspired by the antipodal duality between the three-point FF and the six-gluon amplitude \cite{Dixon:2021tdw}, the $D_4$ symmetry and the similarity of the alphabet (see \cite{Chicherin:2020umh}) suggest possible connections between the four-point FF and the eight-point amplitude.
The antipodal symmetry of the latter was studied recently in \cite{Liu:2022vck}.
With the known two-loop remainder of eight-point MHV amplitude \cite{Caron-Huot:2011zgw, Golden:2013lha},
the four-point FF results provide valuable data to explore such connections.\footnote{After the submission of this paper on arXiv, the symbol of the four-point FF with $q^2\neq0$ was obtained in \cite{Dixon:2022xqh} using symbol bootstrap method, which perfectly confirms our result in the lightlike limit of $q$. A self-antipodal duality for the four-point FF with $q^2\neq0$ was also observed in \cite{Dixon:2022xqh}.}

III) The MI bootstrap construction is based on universal master integrals and can be applied to non-supersymmetric theories.
As in the study of four-point FF of the ${\rm tr}(F^3)$ operator in \cite{Guo:2022pdw},
one can explore the maximally transcendental part for the Higgs-plus-four-gluon amplitude in QCD, which will tell to what extent the maximally transcendental principle \cite{Kotikov:2002ab, Kotikov:2004er, Brandhuber:2012vm, Jin:2018fak, Brandhuber:2017bkg} is applicable for the four-point FF of ${\rm tr}(F^2)$.

IV) The FFs in the strong-coupling limit are computed as minimal-surface areas in $AdS_5$ using Y-system \cite{Alday:2010vh, Maldacena:2010kp, Gao:2013dza, Ouyang:2022sje}.
It would be interesting to extend the construction to the lightlike FFs and compare the OPE prediction with the corresponding Yang-Yang functional at strong coupling.
The lightlike condition and the DDCI are expected to introduce new structures on the Y-system.

V) Finally, we focus on the MHV lightlike FF in this work. It would be worthwhile to consider non-MHV FFs or super-WLs \cite{Mason:2010yk, Caron-Huot:2010ryg, Sever:2011da}, as well as local operators other than the stress-tensor supermultiplet \cite{Engelund:2012re, Brandhuber:2014ica, Wilhelm:2014qua, Nandan:2014oga, Loebbert:2015ova, Brandhuber:2016fni, Loebbert:2016xkw, Banerjee:2016kri, Brandhuber:2018xzk, Lin:2020dyj};
see a recent study of FFOPE involving high-length operators in \cite{Basso:2023bwv}.

\vskip .3cm
{\it Acknowledgments.}
It is a pleasure to thank Qingjun Jin, Guanda Lin, Junbao Wu, Siyuan Zhang, Hua-Xing Zhu, and in particular Song He for discussion and correspondence.
This work is supported in part by the National Natural Science Foundation of China (Grants No.~12175291, 11935013, 12047503, 12247103) and by the CAS under Grants No.~YSBR-101.
We also thank the support of the HPC Cluster of ITP-CAS.


\appendix




\section{One-loop result}\label{app:oneloop}

We summarize the one-loop result in this appendix. The one-loop four-point FF can be given as the cycling summation of the following density functions
\begin{align}
	\hat{{\cal F}}_{4}^{\LL, (1)} = & {\cal G}_{1}^{(1)} + B {\cal G}_{2}^{(1)} \\
    = & \sum_{i=1}^4 \Big[ \hat{{\cal F}}_{4,0}^{(1)}(i,i+1,i+2,i+3) \nonumber \\
    & + \hat{{\cal F}}_{4, \mu}^{(1)}(i,i+1,i+2,i+3) \Big] \,, \nonumber
\end{align}
where
\begin{align}
	\hat{{\cal F}}_{4, 0}^{(1)}(1,2,3,4) = & -\frac{1}{2} I_{\text{Box}}^{(1)}(1,2,3) -\frac{1}{2} I_{\text{Box}}^{(1)}(1,2+3,4) \nonumber\\
	& +I_{\text{Bub}}^{(1)}(1,2,3) \,, \nonumber \\
	\hat{{\cal F}}_{4, \mu}^{(1)}(1,2,3,4) = & \frac{B-1}{2} I_{\text{Pen}}^{(1)}(1,2,3,4) \,.
\end{align}
Here $B$ is the parity-odd factor defined in Eq.\eqref{eq:defB} in the main text, and the one-loop UT MIs are
\begin{align}
	&	I^{(1)}_{\text{Bub}}(1, \ldots, n) = \frac{1-2\epsilon}{\epsilon} \times
	\begin{aligned}
		\includegraphics[scale=0.32]{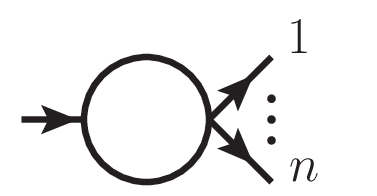}
	\end{aligned} , \\
	&	I^{(1)}_{\text{Box}}(i,j,k) = (s_{ij} s_{jk} -p_j^2 q^2) \times
	\begin{aligned}
		\includegraphics[scale=0.27]{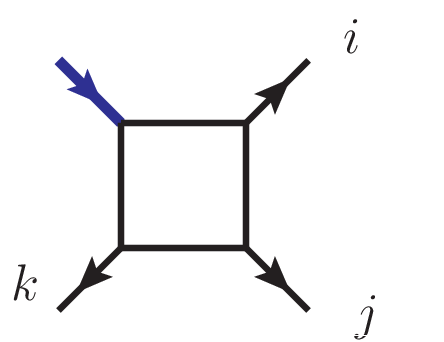}
	\end{aligned} , \\
	&	I^{(1)}_{\text{Pen}}(i,j,k,l) = 4 \mathrm{i} \varepsilon(1234) \times \mu \times
	\begin{aligned}\label{eq:pentMI}
		\includegraphics[scale=0.28]{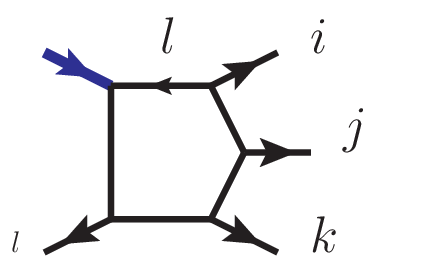}
	\end{aligned} .
\end{align}
In the definition of $I^{(1)}_{\text{Pen}}$, $\mu = l^{-2\epsilon} \cdot l^{-2\epsilon}$ which should be understood as the numerator defined in the pentagon integrand. And $p_j$ in $I^{(1)}_{\text{Box}}(i,j,k)$ can be either a massless or a massive momentum.
We point out that the $4 \mathrm{i} \varepsilon(1234)$ factor is parity odd and the usual scalar pentagon integral is parity even, so here the pentagon MI \eqref{eq:pentMI} is defined with a numerator $4 \mathrm{i} \varepsilon(1234) \times \mu$ which makes it odd.
We note that the one-loop correction proportional to the factor $B$ is
\begin{equation}
{\cal G}_{2}^{(1)} = \sum_{i=1}^4 \frac{1}{2} I_{\text{Pen}}^{(1)}(i,i+1,i+2,i+3)\,,
\end{equation}
which is purely odd and is at ${\cal O}(\epsilon^1)$.

\begin{figure}[ht]
	\centering
	\includegraphics[scale=0.4]{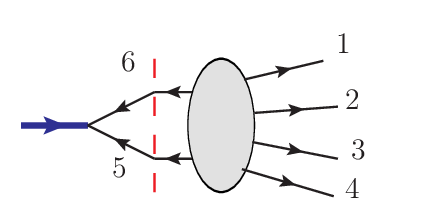}
	\caption{$D$-dimensional cut for the one-loop FF.}
	\label{fig:oneLoopDdimcut}
\end{figure}

The above one-loop result can be obtained via unitarity cuts. In particular, to determine the pentagon contribution, a $D$ dimensional cut is necessary as in Figure~\ref{fig:oneLoopDdimcut}.

Through the BDS ansatz \eqref{eq:BDSansatz}, the one-loop FF provides constraints for the two-loop corrections ${\cal G}_{a}^{(2)}$, which can be written explicitly as
\begin{align}
    {\cal G}_{1}^{(2)}(\epsilon) \big|_{\text{IR}} = & \Big[ \frac{1}{2} \big( {\cal G}_{1}^{(1)}(\epsilon)  \big)^2  + f^{(2)}(\epsilon) {\cal G}_{1}^{(1)}(2\epsilon) \Big] \Big|_{\text{IR}} \,, \nonumber \\
    {\cal G}_{2}^{(2)}(\epsilon) \big|_{\text{IR}} = & {\cal G}_{1}^{(1)}(\epsilon) {\cal G}_{2}^{(1)}(\epsilon)\Big|_{\text{IR}} \,. 
\end{align}
We can see that the IR divergence of ${\cal G}_{2}^{(2)}$ is parity-odd.
Moreover, since ${\cal G}_{2}^{(2)}$ should be able to cancel the odd spurious pole ${4\mathrm{i} \varepsilon(1234)}$ in $B$, this suggests that ${\cal G}_{2}^{(2)}$ is given in terms of purely odd UT integrals.

\section{Numerical check of DDCS}\label{app:numericalDDCI}

We provide the remainder function in terms of pentagon functions, which are defined in \cite{Chicherin:2020oor}. This expression does not manifest the DDCI of the remainder since the pentagon functions depend on Mandelstam variables rather than just three DDCI cross ratios. 
To verify the DDCI of the remainder at the full function level, we preform numerical checks.
We consider two different sets of Mandelstam variables which give the same three cross ratios. 
The two sets of Mandelstam variables are related by a small DDC transformation so that they are in the same kinemtic region. 
We evaluate the integrals and pentagon functions numerically with high precision using \cite{Chicherin:2020oor}. We have also crossed check with DiffExp \cite{Hidding:2020ytt} or AMFlow \cite{Liu:2022chg}, which is applicable to the FF in more general kinematics.
A Sample of numerical results for the two-loop four-point FF and the remainder are shown in the Table~\ref{tab:DDCI}.
One can see that although the FF results are different, the remainder is the same.
We also give the numerical results of the pentagon functions at the three groups of kinematics in the ancillary files.

\begin{table*}[t]
	\centering
	\scalebox{0.9}{
		\begin{tabular}{| c | c | c |}
			\hline
			& \multicolumn{2}{c|}{$\hat{{\cal F}}_4^{\LL, (2)}$} \\ \hline \hline
			$\epsilon^{-4}$ & \multicolumn{2}{c|}{$8$} \\ \hline
			$\epsilon^{-3}$ & $-27.66289379452526 +25.13274122871835 \mathrm{i}$ & $-28.00423336508048 +25.13274122871835 \mathrm{i}$ \\ \hline
			$\epsilon^{-2}$ & $-34.63842478713434 -56.35649295573108 \mathrm{i}$ & $-33.09380826244642 -57.52897381589651 \mathrm{i}$ \\ \hline
			$\epsilon^{-1}$ & $88.97518883469193 - 70.08654798476785 \mathrm{i}$ & $89.11502152100144 - 66.62184859060167 \mathrm{i}$ \\ \hline
			$\epsilon^{ 0}$ & $103.4449187439687 + 73.59353889401273 \mathrm{i}$ & $97.71108801030584 + 74.41653406220033 \mathrm{i}$ \\ \hline \hline
			${\cal R}_{4}^{\LL (2)}$ & \multicolumn{2}{c|}{$-9.67828803915708 + 3.12788590333274 \mathrm{i}$} \\ \hline
		\end{tabular}
	}
	\caption{
		Numerical results of the two-loop four-point lightlike FF:
		the kinematics for the first column are chosen as
		\{$s_{12} = -3$, $s_{13} = 5$, $s_{14} = 6$, $s_{23} = -7$, $s_{24} = -9$, $s_{34} = 8$, $4 \mathrm{i} \varepsilon(1234) = 3\sqrt{399}\mathrm{i}$\},
		and the second column chosen as
		\{$s_{12} = -200/63$, $s_{13} = 52700/10647$, $s_{14} = 1000/169$, $s_{23} = -3500/507$, $s_{24} = -295600/31941$, $s_{34} = 1600/189$, $4 \mathrm{i} \varepsilon(1234) = \sqrt{19/21} \times 10^5/1521 \mathrm{i}$\}.
		\label{tab:DDCI}
	}
\end{table*}

\section{Symbol letters}\label{app:symbolletter}
At the level master integrals, there are $31$ independent symbol letters for all two-loop five-point MIs we use in this paper \cite{Chicherin:2017dob}:
\begin{align}
	& s_{12}\,,\ s_{23}\,,\ s_{34}\,,\ s_{45}\,,\ s_{51} \,, \\
	& s_{13}\,,\ s_{24}\,,\ s_{35}\,,\ s_{14}\,,\ s_{25} \,, \nonumber \\
	& s_{12}+s_{23}\,,\ s_{23}+s_{34}\,,\ s_{34}+s_{45}\,,\ s_{45}+s_{51}\,, \ s_{51}+s_{12}\,, \nonumber \\
	& s_{12}-s_{34}\,,\ s_{23}-s_{45}\,,\ s_{34}-s_{51}\,,\ s_{45}-s_{12}\,, \ s_{51}-s_{12} \,, \nonumber \\
	& s_{12}+s_{24}\,,\ s_{23}+s_{35}\,,\ s_{34}+s_{14} \,,\ s_{45}+s_{25}\,, \ s_{51}+s_{13}\,, \nonumber \\
	& x_{1234}\,,\ x_{1235}\,,\ x_{1245}\,,\ x_{1345} \,,\ x_{2345}\,,\ 4 \mathrm{i} \varepsilon(1234) \,,\nonumber
\end{align}
where $p_5 = -q$ and
\begin{align}
	\label{eq:xijkl-def}
	& x_{ijkl} = \frac{s_{ij} s_{kl} +s_{il} s_{jk} -s_{ik} s_{jl} +4 \mathrm{i} \varepsilon(ijkl)}{s_{ij} s_{kl} +s_{il} s_{jk} -s_{ik} s_{jl} -4 \mathrm{i} \varepsilon(ijkl)} \,.
\end{align}

On the other hand, in the final remainder of the two-loop four-point FF, only nice letters given in Eq.\eqref{eq:alphabet} in the main text are needed to express the remainder symbol as discussed in the main text. If we expand ratio variables $u_i$ in terms of $s_{ij}$, there are only $13$ symbol letters appearing in the remainder:
\begin{align}
	\{ & s_{12} \,, \ s_{23}\,,\ s_{34}\,,\ s_{41}\,,\ s_{1q}\,,\ s_{2q}\,,\ s_{3q}\,,\ s_{4q}\,, \\
	& s_{12}-s_{34}\,,\ s_{23}-s_{41}\,,\ x_{1234}\,,\ x_{123q}\,,\ x_{234q} \} \,. \nonumber
\end{align}

We comment on the appearance of letters at each entry of the two-loop FF remainder:
i) the first entry contains only $s_{i,i+1}$ and $s_{i,i+1,i+2}$ for $i=1,..,4$;
ii) the second entry contains all symbol letters;
iii) the third entry is free from the letters $s_{12}-s_{34}$ and $s_{23}-s_{14}$;
iv) the last entry is free from $s_{12}-s_{34}$, $s_{23}-s_{14}$, $x_{123q}$ and $x_{234q}$.

It is also instructive to write $x$ variables in other forms as
\begin{align}
	x_{1234} = \frac{B+1}{B-1} \,, \ x_{234q} = \frac{B'+1}{B'-1} \,, \ x_{123q} = \frac{B''+1}{B''-1} \,,
\end{align}
where the parity-odd ratio $B$ appears in the one-loop FF in the coefficients of UT MIs, and $B'$ and $B''$ have a similar structure. They can be given in terms of three cross ratios $u_i$ as
\begin{align}
	B^2 = & {\left[-u_2+u_3+u_1 \left(u_2 u_3-1\right)\right]^2}/{Y} \,, \nonumber\\
	{B'}^2 = & {\left[-u_1+u_2+\left(u_1-2\right) u_2 u_3+u_3\right]^2}/{Y} \,, \nonumber \\
	{B''}^2 = & {\left[u_2 \left(u_3-2\right) u_1+u_1+u_2-u_3\right]^2}/{Y} \,, \nonumber \\
	Y = &  \left(u_2-u_3\right)^2+u_1^2 \left(u_2 u_3-1\right)^2 \nonumber \\
	& -2 u_1 \left[u_3 \left(u_2+u_3-4\right) u_2+u_2+u_3\right] \,.
\end{align}
When taking the square-root,  one can fix the signs by matching with \eqref{eq:xijkl-def} as
\begin{align}
	& B = \delta\,\frac{u_2-u_3-u_1 \left(u_2 u_3-1\right)}{\sqrt{Y}},\\
	& B' = \delta\,\frac{-u_1+u_2+\left(u_1-2\right) u_2 u_3+u_3}{\sqrt{Y}}, \\
	& B'' = \delta\,\frac{u_2 \left(u_3-2\right) u_1+u_1+u_2-u_3}{\sqrt{Y}},
\end{align}
where $\delta=\pm1$ depending on the specific kinematic regions.

The four-point FF has cyclic and flip symmetries which in total form a dihedral group $D_4$ (note that the tree-level FF satisfies this symmetry):
\begin{equation}\label{eq:dihedralsym}
\hat{{\cal F}}_4^{(l)} = \hat{{\cal F}}_4^{(l)} |_{p_i \rightarrow p_{i+1}} = \hat{{\cal F}}_4^{(l)} |_{p_i \rightarrow p_{5-i}} \,.
\end{equation}
The nine letters in the final remainder symbol  are closed under the action of the dihedral group $D_4$:
\begin{align}
	& \{u_1, u_2, u_3, 1-u_1, 1-u_2, 1-u_3\} \xlongrightarrow{p_{i} \rightarrow p_{i+1}} \\
	& \{u_2, \frac{1}{u_1}, \frac{1}{u_3}, 1-u_2, -\frac{1-u_1}{u_1}, -\frac{1-u_3}{u_3}\} \,, \nonumber \\
	& \{x_{1234}, x_{234q}, x_{123q}\} \xlongrightarrow{p_{i} \rightarrow p_{i+1}} \{\frac{1}{x_{1234}}, \frac{x_{1234}}{x_{123q}}, x_{234q}\} \,, \nonumber \\
	& \{u_1, u_2, u_3, 1-u_1, 1-u_2, 1-u_3\} \xlongrightarrow{p_{i} \rightarrow p_{5-i}} \nonumber\\
	& \{\frac{1}{u_1}, u_2, \frac{1}{u_3}, \frac{1}{u_3}, -\frac{1-u_1}{u_1}, 1-u_2, -\frac{1-u_3}{u_3}\} \,, \nonumber \\
	& \{x_{1234}, x_{234q}, x_{123q}\} \xlongrightarrow{p_{i} \rightarrow p_{5-i}} \{x_{1234}, \frac{1}{x_{123q}}, \frac{1}{x_{234q}}\} \,.
	\nonumber
\end{align}

\section{One-loop BDS-like functions}\label{app:BDSlike}

As mentioned in the main text, one can introduce ``BDS-like remainder" which is related the BDS remainder as (see also \cite{Dixon:2020bbt, Dixon:2022rse}):
\begin{equation}\label{eq:remainderBDSlike}
	{\cal E}_4 = \exp \bigg[ {\Gamma_{\rm cusp} \over 4} {\cal E}^{(1)}_4 + {\cal R}^\LL_4 \bigg]  \,,
\end{equation}
where the one-loop BDS-like function ${\cal E}^{(1)}_{4}$ can be chosen as the $D_4$-symmetrization of ${\cal W}^{\LL, (1)}_{4}$ given in Eq.\eqref{eq:W1loopOPE} in the main text:
\begin{align}\label{eq:1loopBDSlike1}
	{\cal E}^{(1)}_{4} & = \Big[ \log \left(\frac{(1-u_1) (1-u_2)}{1-u_3}\right) - {1\over2} \log(u_1 u_2) \Big] \\
	& \qquad \times \Big[ \log \left(\frac{(1-u_1) (1-u_2) u_3}{1-u_3}\right) - {1\over2}  \log(u_1 u_2) \Big]. \nonumber
\end{align}
Interestingly, we find that the $1283$ terms of ${\cal S}({\cal R}_4^{\LL, (2)})$ cancel significantly with ${\cal S}(({\cal E}^{(1)}_{4})^2/2)$ such that ${\cal S}({\cal E}^{(2)}_4)$ contains only $456$ terms.
This implies that the BDS-like remainder may have simpler symbol structure comparing to the BDS remainder.
(One should keep in mind that the BDS remainder has nicer collinear property though.)

Note that BDS-like function is not uniquely defined, as one may choose a different one-loop function ${\cal E}^{(1)}_4$ in \eqref{eq:remainderBDSlike}.
For example, another choice is
\begin{align}\label{eq:1loopBDSlike2}
	{\cal E}'^{(1)}_{4} & = \log \left[\frac{(1-u_1) (1-u_2)}{1-u_3}\right]  \log \left[\frac{(1-u_1) (1-u_2) u_3}{(1-u_3) u_1 u_2}\right] \nonumber\\
	& \quad +  { \log(u_1)\log(u_2) - \log(u_1 u_2) \log(u_3) \over 2} .
\end{align}
We comment that different choice will not change the alphabet of the two-loop finite remainder.

\section{FFOPE parametrization}\label{app:OPEparameter}

\begin{figure}[t]
	\centering
	\includegraphics[scale=0.42]{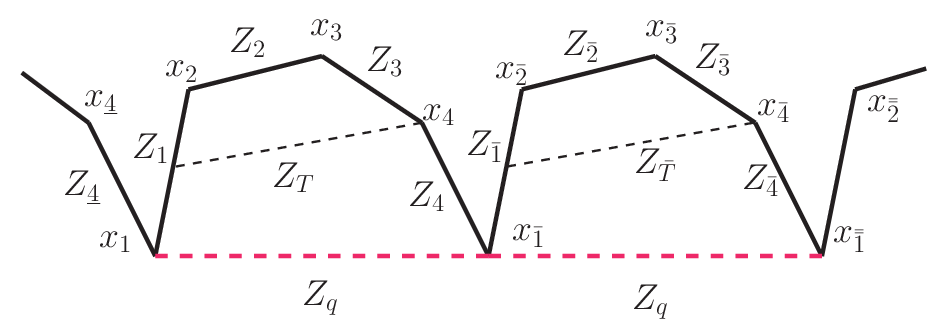}
	\caption{Twistors in OPE.}
	\label{fig:OPEtwistor}
\end{figure}

In this appendix we provide some details on the parametrization of the WL used in the OPE picture.
We define the twistor variables as in Figure~\ref{fig:OPEtwistor}.

The momentum twistors $Z_i$ are related spinor variables in the standard way as
\begin{align}
	Z_{i}^A = (\lambda_i^\alpha, \mu_i^{\dot\alpha}) , \quad \mu_i^{\dot\alpha} = x_i^{\alpha{\dot\alpha}} \cdot \lambda_{i \alpha} =  x_{i+1}^{\alpha{\dot\alpha}} \cdot \lambda_{i \alpha}  .
\end{align}
Here the spinor $\lambda_i$ is related to the on-shell momentum as in the standard spinor helicity formalism:
\begin{equation}
	p_{i,\mu} \rightarrow p_{i,\alpha\dot\alpha} = p_{i,\mu} \sigma^\mu_{\alpha\dot\alpha} =  \lambda_{i,\alpha} \tilde\lambda_{i,\dot\alpha} \,.
\end{equation}

An explicit choice of the twistor variables can be chosen as follows.
Similar to the paper \cite{Basso:2013aha}, we first set the following twistors as constants:
\begin{align}
	Z_{q}=\{1,0,0,0\}\,,\quad
	Z_4&=\{1,-1,-1,1\}\,,\\
	Z_{\underline{4}}=\{1,-1,0,0\}\,,\quad
	Z_{\bar 4}&=\{1,-1,-2,2\}\,,\\
	Z_{T}=\{1,1,-1,-1\}\,,\quad
	Z_{\bar T}&=\{1,1,0,-2\}\,,\\
	Z_1=\{0,1,0,0\}\,,\quad
	Z_{\bar{1}}&=\{0,1,1,-1\}\,.
\end{align}
Then we define the remaining twistors that depend on the $\{ \tau, \sigma, \phi\}$  via transformation
\begin{align}
	Z_2=\{2,2,-1,-1\}.M\,,\quad
	Z_3&=\{0,2,0,-2\}.M\,,\\
	Z_{\bar{2}}=\{2,2,1,-3\}.\bar{M}\,,\quad
	Z_{\bar{3}}&=\{0,2,2,-4\}.\bar{M}\,,
\end{align}
where the transformation matrices $M$ and $\bar{M}$ are
\begin{widetext}
	\begin{align}
		&M=\left(
		\begin{array}{cccc}
			\mathrm{e}^{-\tau -\frac{\mathrm{i} \phi }{2}} & 0 & 0 & 0 \\
			0 & \mathrm{e}^{\sigma +\frac{\mathrm{i} \phi }{2}} & 0 & 0 \\
			-\frac{1}{2} \mathrm{e}^{-\sigma +\frac{\mathrm{i} \phi }{2}}+\mathrm{e}^{-\tau -\frac{\mathrm{i} \phi }{2}}-\frac{1}{2} \mathrm{e}^{\tau -\frac{\mathrm{i} \phi }{2}} & \frac{1}{2} \mathrm{e}^{-\sigma +\frac{\mathrm{i} \phi }{2}}-\frac{1}{2} \mathrm{e}^{\tau -\frac{\mathrm{i} \phi }{2}} & \frac{1}{2} \mathrm{e}^{-\sigma +\frac{\mathrm{i} \phi }{2}}+\frac{1}{2} \mathrm{e}^{\tau -\frac{\mathrm{i} \phi }{2}} & \frac{1}{2} \mathrm{e}^{\tau -\frac{\mathrm{i} \phi }{2}}-\frac{1}{2} \mathrm{e}^{-\sigma +\frac{\mathrm{i} \phi }{2}} \\
			\frac{1}{2} \mathrm{e}^{-\sigma +\frac{\mathrm{i} \phi }{2}}-\frac{1}{2} \mathrm{e}^{\tau -\frac{\mathrm{i} \phi }{2}} & -\frac{1}{2} \mathrm{e}^{-\sigma +\frac{\mathrm{i} \phi }{2}}+\mathrm{e}^{\sigma +\frac{\mathrm{i} \phi }{2}}-\frac{1}{2} \mathrm{e}^{\tau -\frac{\mathrm{i} \phi }{2}} & \frac{1}{2} \mathrm{e}^{\tau -\frac{\mathrm{i} \phi }{2}}-\frac{1}{2} \mathrm{e}^{-\sigma +\frac{\mathrm{i} \phi }{2}} & \frac{1}{2} \mathrm{e}^{-\sigma +\frac{\mathrm{i} \phi }{2}}+\frac{1}{2} \mathrm{e}^{\tau -\frac{\mathrm{i} \phi }{2}} \\
		\end{array}
		\right)\,,\\
		&\bar{M}=\left(
		\begin{array}{cccc}
			\mathrm{e}^{-\tau -\frac{\mathrm{i} \phi }{2}} & 0 & 0 & 0 \\
			\mathrm{e}^{-\sigma +\frac{\mathrm{i} \phi }{2}}-\mathrm{e}^{-\tau -\frac{\mathrm{i} \phi }{2}} & 2 \mathrm{e}^{\sigma +\frac{\mathrm{i} \phi }{2}}-\mathrm{e}^{-\sigma +\frac{\mathrm{i} \phi }{2}} & 2 \mathrm{e}^{\sigma +\frac{\mathrm{i} \phi }{2}}-2 \mathrm{e}^{-\sigma +\frac{\mathrm{i} \phi }{2}} & 2 \mathrm{e}^{-\sigma +\frac{\mathrm{i} \phi }{2}}-2 \mathrm{e}^{\sigma +\frac{\mathrm{i} \phi }{2}} \\
			-\frac{1}{2} \mathrm{e}^{-\sigma +\frac{\mathrm{i} \phi }{2}}+\mathrm{e}^{-\tau -\frac{\mathrm{i} \phi }{2}}-\frac{1}{2} \mathrm{e}^{\tau -\frac{\mathrm{i} \phi }{2}} & \frac{1}{2} \mathrm{e}^{-\sigma +\frac{\mathrm{i} \phi }{2}}-\frac{1}{2} \mathrm{e}^{\tau -\frac{\mathrm{i} \phi }{2}} & \mathrm{e}^{-\sigma +\frac{\mathrm{i} \phi }{2}} & \mathrm{e}^{\tau -\frac{\mathrm{i} \phi }{2}}-\mathrm{e}^{-\sigma +\frac{\mathrm{i} \phi }{2}} \\
			\frac{1}{2} \mathrm{e}^{-\sigma +\frac{\mathrm{i} \phi }{2}}-\frac{1}{2} \mathrm{e}^{\tau -\frac{\mathrm{i} \phi }{2}} & -\frac{1}{2} \mathrm{e}^{-\sigma +\frac{\mathrm{i} \phi }{2}}+\mathrm{e}^{\sigma +\frac{\mathrm{i} \phi }{2}}-\frac{1}{2} \mathrm{e}^{\tau -\frac{\mathrm{i} \phi }{2}} & \mathrm{e}^{\sigma +\frac{\mathrm{i} \phi }{2}}-\mathrm{e}^{-\sigma +\frac{\mathrm{i} \phi }{2}} & \mathrm{e}^{-\sigma +\frac{\mathrm{i} \phi }{2}}-\mathrm{e}^{\sigma +\frac{\mathrm{i} \phi }{2}}+\mathrm{e}^{\tau -\frac{\mathrm{i} \phi }{2}} \\
		\end{array}
		\right)\,.
	\end{align}
\end{widetext}

One can check that the above formulae satisfy the self-consistence condition
\begin{align}
	P(Z_iM)=P(Z_i)P(M)\,,
\end{align}
where $P$ denotes the periodic translation
\begin{align}
	P(Z_i)\equiv Z_{\bar{i}}=Z_i+ \{0,0,q^{\alpha\dot{\alpha}}\lambda_{i\alpha}\}\,,
\end{align}
and $P(M)\equiv \bar{M}$.

The OPE parameters $\{ \tau, \sigma, \phi\}$ are related to the three ratios $u_i$ in Eq.\eqref{eq:3DDCIratios} in the main text as follows:
\begin{align}
	u_1& = \frac{x^2_{13}}{x^2_{3{\bar 1}}} = \frac{\langle 4 \bar{1}\rangle  \langle \underline{4} 1 2 3\rangle }{\langle \underline{4} 1\rangle  \langle 2 3 4 \bar{1}\rangle } =\mathrm{e}^{-2 \sigma } \,, \\
	u_2& = \frac{x^2_{24}}{x^2_{4{\bar 2}}} = \frac{\langle\bar{1} \bar{2}\rangle  \langle 1 2 3 4\rangle }{\langle 1 2\rangle  \langle 3 4 \bar{1} \bar{2}\rangle } =\mathrm{e}^{-2 \tau }\,,  \\
	u_3& = \frac{x^2_{14} x^2_{3{\bar 2}}}{x^2_{2{\bar 1}} x^2_{4{\bar 3}}} = \frac{\langle 1 2\rangle  \langle 4 \bar{1}\rangle  \langle \bar{2} \bar{3}\rangle  \langle \underline{4} 1 3 4\rangle  \langle 2 3 \bar{1} \bar{2}\rangle }{\langle \underline{4} 1\rangle  \langle 2 3\rangle  \langle\bar{1} \bar{2}\rangle  \langle 1 2 4 \bar{1}\rangle  \langle 3 4 \bar{2} \bar{3}\rangle } \notag\\&=\frac{\cosh (\sigma -\tau )+\cos (\phi )}{\cosh (\sigma +\tau )+\cos (\phi )} \,.
\end{align}
The angle brackets represent the contract of spinors or momentum twistors via Levi-Civita tensors as:
\begin{align}
	&\langle i j\rangle := \epsilon_{\alpha\beta}\lambda_i^\alpha \lambda_j^{\beta} \ , \\
	&\langle i j k l\rangle := \epsilon_{ABCD}Z_i^A Z_j^B Z_k^C Z_l^D \,.
\end{align}

The $D_4$ symmetry acting on the OPE parameters gives
\begin{align}
	&\{\tau, \sigma\}\ \xlongrightarrow[]{\mbox{$p_i \rightarrow p_{i+1}$}} \ \{-\sigma, \tau\} \,,
	\nonumber\\
	&\{\tau, \sigma\}\ \xlongrightarrow[]{\mbox{$p_i \rightarrow p_{5-i}$}} \ \{-\tau, -\sigma\} \,.
\end{align}
This implies that the BDS remainder satisfies
\begin{equation}
	{\cal R}^\LL_4(\tau, \sigma, \phi) = {\cal R}^\LL_4(\sigma, \tau, \phi) = {\cal R}^\LL_4(\tau, -\sigma, \phi) \,.
\end{equation}

Following the above parametrization of the momentum twistors, we perform the calculation (such as the one-loop result \eqref{eq:W1loopOPE}) within the kinematics region
\begin{align}\label{eq:opeMandelstamregion}
	& \{s_{12}>0\,,\ s_{23}<0\,,\ s_{34}>0\,,\ s_{14}<0\,,\nonumber \\
	& s_{123}<0\,,\ s_{124}<0\,,\ s_{234}>0\,,\ s_{134}>0 \}\,.
\end{align}

\section{Some details on FFOPE}\label{app:OPEdetail}

We provide more details on the OPE computation.
The four-point finite ratio defined in Eq.\eqref{eq:W4ratio} in the main text can be given in a more concrete form as
\begin{align}
	{\cal W}^\LL_{4}  = \sum_{\bf a} \int d {\bf u} \, \mathrm{e}^{-E_{\bf a}({\bf u}) \tau +\mathrm{i} p_{\bf a}({\bf u}) \sigma +\mathrm{i} m_{\bf a} \phi} \, {\mathbb P}_{{\bf a}}(0 | {\bf u}) \, {\mathbb F}_{\bar{\bf a}}^\LL(\bar{\bf u}) \,,
\end{align}
where the GKP eigenstates $\psi$ are represented by ${\bf a}$ and their Bethe rapidities ${\bf u}$, the integral measure is
\begin{equation}
	d{\bf u} = \prod_{i=1}^N {d u_i \over 2\pi}  \mu_{a_i}(u_i) \,.
\end{equation}

\subsubsection*{Single gluonic excitation}

The leading contribution of the OPE comes from the lightest state:  gluonic excitation states $F$ and ${\bar F}$, similar to the six-gluon case (rather than the FF with $q^2\neq0$ case), and this leads to
\begin{align}
	{\cal W}^\LL_{4}  = 1 + 2 \cos(\phi) h_0(\tau, \sigma) + ... \,,
\end{align}
where
\begin{equation}
	h_0(\tau, \sigma) = \int {d u \over 2\pi} \, \mu_F(u) \, \mathrm{e}^{-E(u) \tau +\mathrm{i} p(u) \sigma} \, {\mathbb P}(0 | u) \, {\mathbb F}_{\scriptscriptstyle{F}}^\LL(u) .
\end{equation}
One can normalize the pentagon transition ${\mathbb P}(0 | u)$ to be equal one \cite{Basso:2013aha}, and
\begin{equation}
	\mu_{F}(u) = -\frac{\pi g^2}{(u^2 +\frac{1}{4}) \cosh(\pi u)} (1 + {\cal O}(g^2)) \,.
\end{equation}

To consider the dependence of $\tau$, one uses $E(u) = 1 + \gamma(u)$ and obtains
\begin{align}
	h_0(\tau, \sigma) = & \mathrm{e}^{-\tau} \int {d u \over 2\pi} \mathrm{e}^{-\gamma(u) \tau +\mathrm{i} p(u) \sigma} \nonumber\\
	& \times {-\pi g^2 (1 + {\cal O}(g^2)) \over (u^2 +\frac{1}{4}) \cosh(\pi u)} \ {\mathbb F}_{\scriptscriptstyle{F}}^\LL(u)  \,. \label{eq:h0inapp}
\end{align}

Let us study the expansion of \eqref{eq:h0inapp} in $g$ in detail. We note that $\gamma(u)$ is at order $g^2$ and $p(u) = 2u + {\cal O}(g^2)$.
If ${\mathbb F}_{\scriptscriptstyle{F}}^\LL(u)$ starts at $g^2$ order:
\begin{equation}
	{\mathbb F}_{\scriptscriptstyle{F}}^\LL(u) = g^2 y_{\scriptscriptstyle{F}}(u) + {\cal O}(g^4) \,,
\end{equation}
the leading terms $\mathrm{e}^{-\tau} \tau^{\ell-1}$ are zero to all order.
As discussed in the main text, this is what one would expect from the one- and two-loop results, and this simple observation leads to all-order predictions:
\begin{equation}
	{\cal W}^{\LL, (\ell)}_{4} |_{\tau^{\ell-1}\mathrm{e}^{-\tau} } = 0 \,, \quad \textrm{for all} \ \ell \,.
\end{equation}
At two-loop order, one has the leading order expansion at $\mathrm{e}^{- \tau }$ as
\begin{equation}
	{\cal W}^{\LL, (2)}_{4}  = 2 \mathrm{e}^{-\tau} \cos(\phi) h_0^{(2)}(\sigma) + {\cal O}(\mathrm{e}^{-2\tau}) ,
\end{equation}
with
\begin{equation}\label{eq:h0FT1}
	h_0^{(2)}(\sigma) = \int {d u \over 2\pi} \, {- \pi \, y_{\scriptscriptstyle{F}}(u)  \over  (u^2 +\frac{1}{4}) \cosh(\pi u)} \, \mathrm{e}^{\mathrm{i} 2u \sigma} \,.
\end{equation}
From the two-loop FF result, one can extract $h_0^{(2)}(\sigma)$ as
\begin{align}
	h_{0}^{(2)}(\sigma) = & 4 \mathrm{e}^\sigma\big[ -{\rm Li}_3(\mathrm{e}^{-2\sigma}) + {\rm Li}_2(1-\mathrm{e}^{-2\sigma})  \\
	& \qquad -\sigma {\rm Li}_2(\mathrm{e}^{-2\sigma}) -\zeta_2 \big] +(\sigma \rightarrow -\sigma) \,. \nonumber
\end{align}
An inverse Fourier transformation of \eqref{eq:h0FT1} gives
\begin{equation}
	y_{\scriptscriptstyle{F}}(u) = \frac{2\pi}{(u^2 +\frac{1}{4}) \cosh(\pi u)} = -2\mu_F(u)\big|_{g^2} \,,
\end{equation}
where $\mu_F(u)$ is the measure of the gluonic excitation~\cite{Basso:2013vsa}.
With this result, one can make another all-order prediction as given in Eq.\eqref{eq:OPEprediction2} in the main text.

\subsubsection*{Multiple-particle excitations}

Below we comment further on the multiple particle excitations. Since the one-loop OPE has non-zero terms of $\mathrm{e}^{- m \tau}$ with $m\geq2$, the transition corresponding to multiple particles should start at ${\cal O}(g^0)$ order.

In particular, the order of $\mathrm{e}^{-2\tau}$ comes from two-particle and bound-state excitations. Similar to the six-point amplitude, we have
\begin{align}
	{\cal W}^{\LL, (\ell)}_{4}\big|_{\mathrm{e}^{-2\tau}}  = \sum_{k=1}^{\ell-1} \tau^{k-1} {\hat h}_{k}^{(\ell)}(\sigma, \phi) \,.
\end{align}
For the FF results up to two loops, we obtain
\begin{align}
	{\hat h}_{2}^{(2)}(\sigma,\phi) = & 8(2 +\cos(2\phi)) \,, \\
	{\hat h}_{1}^{(2)}(\sigma,\phi) = & 2 (4 \sigma^2 +5 \zeta_2 +3)(1+\cos(2\phi)) +14 \nonumber \\
	& +2 \cos(2\phi) \big[\mathrm{e}^{2 \sigma } \big( 2 \sigma {\rm Li}_2(\mathrm{e}^{-2 \sigma}) -{\rm Li}_2(1-\mathrm{e}^{-2 \sigma}) \nonumber \\
	& +2 {\rm Li}_3(\mathrm{e}^{-2 \sigma }) +\zeta_2\big) +(\sigma \rightarrow -\sigma) \big]  \,. \nonumber
\end{align}
We point out that the parts proportional to different $\cos(m\phi)$ are from the contributions carried $U(1)$ charge $\pm m$, respectively.

Finally, we mention that our discussion for the four-point lightlike FF is stratight to generalize to higher point cases. For an $n$-point light FF, it can be decomposed into a three-sided null-wrapped polygon and $n-3$ pentagons:
\begin{align}\label{eq:WLLngeneralOPE}
	{\cal W}^\LL_{n} & = \sum_{\psi_1, .., \psi_{n-3}} \mathrm{e}^{\sum_j (-E_j \tau_j +\mathrm{i} p_j \sigma_j +\mathrm{i} m_j \phi_j)} \nonumber\\
	& \times {\mathbb P}(0 | \psi_1) \ldots {\mathbb P}(\psi_{n-4} | \psi_{n-3}) {\mathbb F}^\LL(\psi_{n-3}) \,.
\end{align}
The $3n-9$ independent conformal ratios are related to the symmetries of $n-3$ null squares in the decomposition, parametrized by $\{\tau_i, \sigma_i, \phi_i\}$.

\section{Comparing different OPE formalisms}\label{app:comparedifferentOPE}

Here we compare the lightlike FFOPE with the previous known formalism in more detail.

\begin{figure}[t]
	\centering
	\includegraphics[scale=0.48]{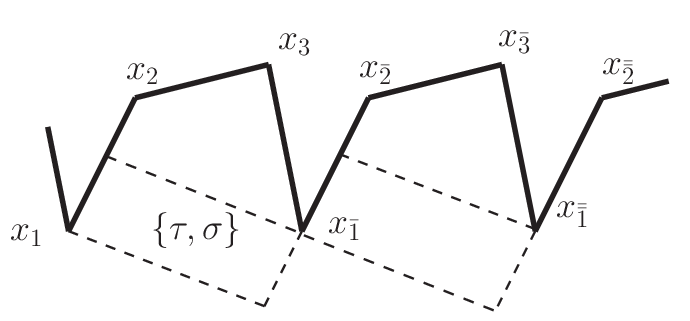}
	\caption{Dual periodic WL for the three-point FF ${\cal W}_{{\cal F}_{3}}$.}
	\label{fig:PWLF3}
\end{figure}

Firstly, we consider the three-point FF with $q^2\neq0$, denoted as ${\cal W}_{{\cal F}_{3}}$, which is the first non-trivial case in the FFOPE \cite{Sever:2020jjx}.
The dual periodic WL picture is shown in Figure~\ref{fig:PWLF3}.
Since $q^2\neq0$, there are two independent ratio variables which can be chosen as \cite{Sever:2020jjx}
\begin{align}
	u_1&={x^2_{{\bar 1}{\bar {\bar 1}}} x^2_{2{\bar 2}} \over  x^2_{2{\bar 1}} x^2_{{\bar 2}{\bar {\bar 1}}}}=(1+\mathrm{e}^{2 \tau})^2\,,\\
	u_2&={x_{1\bar{1}}^2x_{3\bar{3}}^2\over  x_{13}^2x_{\bar{1}\bar{3}}^2}=(1+\mathrm{e}^{-2 \tau}+\mathrm{e}^{2 \sigma})^2\,.
\end{align}
In particular there is no $\phi$ variable due to the $U(1)_\phi$ symmetry, which is different from the lightlike FFOPE.
Furthermore, related to the $U(1)_\phi$ symmetry, the leading contribution to the OPE of ${\cal W}_{{\cal F}_{3}}$ comes from two-particle singlet states \cite{Sever:2020jjx}. Correspondingly, only even powers of $\mathrm{e}^{-\tau}$ can appear in the large $\tau$ expansion of ${\cal W}_{{\cal F}_{3}}$ at any loop order.

\begin{figure}[b]
	\centering
	\includegraphics[scale=0.48]{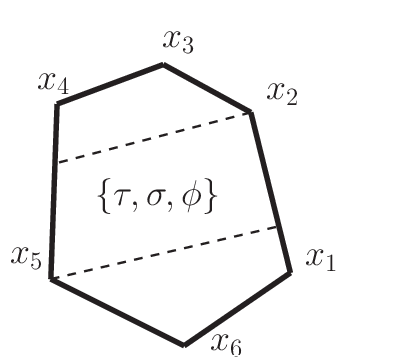}
	\caption{The WL dual to six-point amplitude ${\cal W}_{{\cal A}_{6}}$.}
	\label{fig:WLA6}
\end{figure}

Next, regarding the excitations in the OPE expansion, the lightlike FFOPE we develop is more similar to the amplitude case ${\cal W}_{{\cal A}_{6}}$  \cite{Basso:2013vsa}.
Both the ${\cal W}^{\rm LL}_4$ and ${\cal W}_{{\cal A}_{6}}$ have three cross ratios. Moreover, their leading contributions are both the single gluon excitation. On the other hand, the comformal ratio variables and their parametrization are very different from each other.
We recall the parametrization of ${\cal W}_{{\cal A}_{6}}$ given in \cite{Basso:2013aha}:
\begin{align}
	u_3&={x^2_{26} x^2_{35}\over  x^2_{25} x^2_{36}}=\frac{1}{1+\mathrm{e}^{2 \sigma}+2 \mathrm{e}^{\sigma-\tau} \cos (\phi)+\mathrm{e}^{-2 \tau}}\,,\\
	u_2&={x^2_{15} x^2_{24}\over  x^2_{14} x^2_{25}}=\frac{1}{2} \mathrm{e}^{-\tau} \operatorname{sech}(\tau)\,,\\
	u_1&={x^2_{13} x^2_{46}\over  x^2_{14} x^2_{36}}=\mathrm{e}^{2 \sigma+2 \tau} u_2 u_3\,,
\end{align}
where the WL configuration is shown in Figure~\ref{fig:WLA6}.
One can compare this with the parametrization for the lightlike FF in Figuer~\ref{fig:OPEparametrization}:
\begin{align}
u_1 &= \frac{x^2_{13}}{x^2_{3\bar{1}}} = \mathrm{e}^{-2\sigma}\,, \\
u_2 &= \frac{x^2_{24}}{x^2_{4\bar{2}}} = \mathrm{e}^{-2\tau} \,, \\
u_3 &= \frac{x^2_{14} x^2_{3\bar{2}}}{x^2_{2\bar{1}} x^2_{4\bar{3}}} = \frac{ \cosh(\sigma-\tau) + \cos(\phi)}{\cosh(\sigma+\tau) + \cos(\phi)} \,,
\end{align}
which is clearly very different.

\begin{figure}[t]
	\centering
	\includegraphics[scale=0.48]{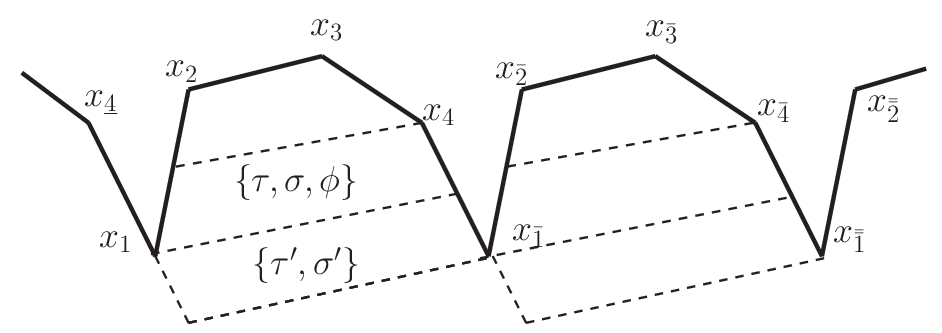}
	\caption{Periodic WL of ${\cal W}_{{\cal F}_{4}}$ with $q^2\neq0$.}
	\label{fig:OPEtwistorF4}
\end{figure}

Finally, one may wonder if the lightlike FFOPE can be obtained from the FFOPE in \cite{Sever:2020jjx} by taking a lightlike limit. For example, the regularized ratio function ${\cal W}^{\LL}_{4}$ might be formally obtained as a limit:
\begin{equation}\label{eq:limitW4}
	{\cal W}^{\LL}_{4} \sim {{\cal W}_{{\cal F}_{4}} \over {\cal W}_{{\cal F}_{3}}} \Big|_{q^2\rightarrow0} \,,
\end{equation}
where ${\cal W}_{{\cal F}_{n}}$ on the RHS are FF with $q^2\neq0$.
However, taking such a limit is highly non-trivial and probably not feasible in practice.
The limit as in \eqref{eq:limitW4} would require cancellations of singular terms, and since the OPE formalism is at non-perturbative level, this could be a formidable task.
Moreover, there are five ratios in the case of ${\cal W}_{{\cal F}_{4}}$ (see Figure~\ref{fig:OPEtwistorF4}), but only three ratios survive in the lightlike FF ${\cal W}^{\LL}_{4}$.
While the lightlike condition  would reduce only one degree of freedom, the hidden dual conformal symmetry of the lightlike FF reduces another one.
It is also non-trivial to make the emergence of the symmetry manifest at non-perturbative level.
Therefore, it should be reasonable that one should apply the lightlike condition from the very beginning to formulate the OPE, and the WL with lightlike periodic should be taken as a genuin new configuration.

\section{Towards high-loop symbol bootstrap}
\label{app:symbolbootstrap}

\noindent

Based on the simple symbol alphabet \eqref{eq:alphabet}, and combined with the OPE prediction, it is natural to bootstrap the four-point FF to higher loops like the three-point case \cite{Dixon:2020bbt, Dixon:2022rse}.
As a preliminary study, we consider the construction of the remainder using the symbol bootstrap method up to three loops.

We start with an ansatz with only the letters in \eqref{eq:alphabet} and no $x_{ijkl}$ in the first entry.
We mention that this is conjectural and we cannot rule out the possibility that new letters may appear at three- or higher-loop level.
Next, we impose the $D_4$ symmetry and the integrability condition for the symbol.
Furthermore, we require the symbol to be zero in the collinear limit.
We then apply two ``minimal" entry conditions:
1) the first-entry must be either $s_{i,i+1}$ or $s_{i,i+1,i+2}$ (the branch cut condition), and
2) the first $L$ entries can not all be the same $s_{i,i+1}$  (the $L^{\rm th}$ discontinuity condition) \cite{Dixon:2020bbt, Gaiotto:2011dt}.
Finally, we impose the OPE constraints at $\mathrm{e}^{-\tau}$ order.
We summarize the constraints and the change of free parameters in Table~\ref{tab:symbolbootstrap}.

\begin{table}[ht]
	\centering
	\vskip .1 cm
	\begin{tabular}{| c | c | c |}
		\hline
		Constraints				&  \multicolumn{2}{c|}{Parameters left}   \\ \hline \hline
		Starting ansatz   		&  4374 & 354294  	\\ \hline
		$D_4$ symmetry   		&  561 	& 44409		\\ \hline
		Integrability			&  72 	& 1056   	\\ \hline
		Collinear limit 		&  56 	& 992  		\\ \hline
		Branch cut condition	&  21 	& 295  		\\ \hline
		$L^{\rm th}$ discontinuity condition		&  8 	& 206  	\\ \hline
		FFOPE (leading $\mathrm{e}^{-\tau} \tau^{L-1}$) 		&  5 	& 193  	\\ \hline
		FFOPE ($\mathrm{e}^{-\tau} \tau^\ell$, $\ell<L-1$) 	&  1 	& 126  	\\ \hline
		\emph{Loops} 			&  2-loop 	& 3-loop  	\\ \hline
	\end{tabular}
	\caption{Symbol bootstrap at 2 and 3 loops.
		\label{tab:symbolbootstrap}}
\end{table}

One can see that the OPE prediction at the $\mathrm{e}^{-\tau}$ order, which are only from the single gluonic excitation, can provide strong constraints.
Note that it would important to explore other entry conditions for the letters, and with a better understanding of Steinmann-like relations and further entry properties as in \cite{Dixon:2020bbt, Dixon:2022rse}, it is highly promising that the three-loop results can be uniquely fixed.

%


\begin{thebibliography}{106}%
	\makeatletter
	\providecommand \@ifxundefined [1]{%
	 \@ifx{#1\undefined}
	}%
	\providecommand \@ifnum [1]{%
	 \ifnum #1\expandafter \@firstoftwo
	 \else \expandafter \@secondoftwo
	 \fi
	}%
	\providecommand \@ifx [1]{%
	 \ifx #1\expandafter \@firstoftwo
	 \else \expandafter \@secondoftwo
	 \fi
	}%
	\providecommand \natexlab [1]{#1}%
	\providecommand \enquote  [1]{``#1''}%
	\providecommand \bibnamefont  [1]{#1}%
	\providecommand \bibfnamefont [1]{#1}%
	\providecommand \citenamefont [1]{#1}%
	\providecommand \href@noop [0]{\@secondoftwo}%
	\providecommand \href [0]{\begingroup \@sanitize@url \@href}%
	\providecommand \@href[1]{\@@startlink{#1}\@@href}%
	\providecommand \@@href[1]{\endgroup#1\@@endlink}%
	\providecommand \@sanitize@url [0]{\catcode `\\12\catcode `\$12\catcode `\&12\catcode `\#12\catcode `\^12\catcode `\_12\catcode `\%12\relax}%
	\providecommand \@@startlink[1]{}%
	\providecommand \@@endlink[0]{}%
	\providecommand \url  [0]{\begingroup\@sanitize@url \@url }%
	\providecommand \@url [1]{\endgroup\@href {#1}{\urlprefix }}%
	\providecommand \urlprefix  [0]{URL }%
	\providecommand \Eprint [0]{\href }%
	\providecommand \doibase [0]{http://dx.doi.org/}%
	\providecommand \selectlanguage [0]{\@gobble}%
	\providecommand \bibinfo  [0]{\@secondoftwo}%
	\providecommand \bibfield  [0]{\@secondoftwo}%
	\providecommand \translation [1]{[#1]}%
	\providecommand \BibitemOpen [0]{}%
	\providecommand \bibitemStop [0]{}%
	\providecommand \bibitemNoStop [0]{.\EOS\space}%
	\providecommand \EOS [0]{\spacefactor3000\relax}%
	\providecommand \BibitemShut  [1]{\csname bibitem#1\endcsname}%
	\let\auto@bib@innerbib\@empty
	\bibitem [{\citenamefont {Alday}\ and\ \citenamefont {Roiban}(2008)}]{Alday:2008yw}%
	  \BibitemOpen
	  \bibfield  {author} {\bibinfo {author} {\bibfnamefont {L.~F.}\ \bibnamefont {Alday}}\ and\ \bibinfo {author} {\bibfnamefont {R.}~\bibnamefont {Roiban}},\ }\href {\doibase 10.1016/j.physrep.2008.08.002} {\bibfield  {journal} {\bibinfo  {journal} {Phys. Rept.}\ }\textbf {\bibinfo {volume} {468}},\ \bibinfo {pages} {153} (\bibinfo {year} {2008})},\ \Eprint {http://arxiv.org/abs/0807.1889} {arXiv:0807.1889 [hep-th]} \BibitemShut {NoStop}%
	\bibitem [{\citenamefont {Arkani-Hamed}\ \emph {et~al.}(2011)\citenamefont {Arkani-Hamed}, \citenamefont {Bourjaily}, \citenamefont {Cachazo}, \citenamefont {Caron-Huot},\ and\ \citenamefont {Trnka}}]{Arkani-Hamed:2010zjl}%
	  \BibitemOpen
	  \bibfield  {author} {\bibinfo {author} {\bibfnamefont {N.}~\bibnamefont {Arkani-Hamed}}, \bibinfo {author} {\bibfnamefont {J.~L.}\ \bibnamefont {Bourjaily}}, \bibinfo {author} {\bibfnamefont {F.}~\bibnamefont {Cachazo}}, \bibinfo {author} {\bibfnamefont {S.}~\bibnamefont {Caron-Huot}}, \ and\ \bibinfo {author} {\bibfnamefont {J.}~\bibnamefont {Trnka}},\ }\href {\doibase 10.1007/JHEP01(2011)041} {\bibfield  {journal} {\bibinfo  {journal} {JHEP}\ }\textbf {\bibinfo {volume} {01}},\ \bibinfo {pages} {041} (\bibinfo {year} {2011})},\ \Eprint {http://arxiv.org/abs/1008.2958} {arXiv:1008.2958 [hep-th]} \BibitemShut {NoStop}%
	\bibitem [{\citenamefont {Beisert}\ \emph {et~al.}(2012)\citenamefont {Beisert} \emph {et~al.}}]{Beisert:2010jr}%
	  \BibitemOpen
	  \bibfield  {author} {\bibinfo {author} {\bibfnamefont {N.}~\bibnamefont {Beisert}} \emph {et~al.},\ }\href {\doibase 10.1007/s11005-011-0529-2} {\bibfield  {journal} {\bibinfo  {journal} {Lett. Math. Phys.}\ }\textbf {\bibinfo {volume} {99}},\ \bibinfo {pages} {3} (\bibinfo {year} {2012})},\ \Eprint {http://arxiv.org/abs/1012.3982} {arXiv:1012.3982 [hep-th]} \BibitemShut {NoStop}%
	\bibitem [{\citenamefont {Dixon}(2014)}]{Dixon:2013uaa}%
	  \BibitemOpen
	  \bibfield  {author} {\bibinfo {author} {\bibfnamefont {L.~J.}\ \bibnamefont {Dixon}}\ }(\bibinfo {year} {2014})\ pp.\ \bibinfo {pages} {31--67},\ \Eprint {http://arxiv.org/abs/1310.5353} {arXiv:1310.5353 [hep-ph]} \BibitemShut {NoStop}%
	\bibitem [{\citenamefont {Elvang}\ and\ \citenamefont {Huang}(2013)}]{Elvang:2013cua}%
	  \BibitemOpen
	  \bibfield  {author} {\bibinfo {author} {\bibfnamefont {H.}~\bibnamefont {Elvang}}\ and\ \bibinfo {author} {\bibfnamefont {Y.-t.}\ \bibnamefont {Huang}},\ }\href@noop {} {\  (\bibinfo {year} {2013})},\ \Eprint {http://arxiv.org/abs/1308.1697} {arXiv:1308.1697 [hep-th]} \BibitemShut {NoStop}%
	\bibitem [{\citenamefont {Henn}\ and\ \citenamefont {Plefka}(2014)}]{Henn:2014yza}%
	  \BibitemOpen
	  \bibfield  {author} {\bibinfo {author} {\bibfnamefont {J.~M.}\ \bibnamefont {Henn}}\ and\ \bibinfo {author} {\bibfnamefont {J.~C.}\ \bibnamefont {Plefka}},\ }\href {\doibase 10.1007/978-3-642-54022-6} {\emph {\bibinfo {title} {{Scattering Amplitudes in Gauge Theories}}}},\ Vol.\ \bibinfo {volume} {883}\ (\bibinfo  {publisher} {Springer},\ \bibinfo {address} {Berlin},\ \bibinfo {year} {2014})\BibitemShut {NoStop}%
	\bibitem [{\citenamefont {Travaglini}\ \emph {et~al.}(2022)\citenamefont {Travaglini} \emph {et~al.}}]{Travaglini:2022uwo}%
	  \BibitemOpen
	  \bibfield  {author} {\bibinfo {author} {\bibfnamefont {G.}~\bibnamefont {Travaglini}} \emph {et~al.},\ }\href {\doibase 10.1088/1751-8121/ac8380} {\bibfield  {journal} {\bibinfo  {journal} {J. Phys. A}\ }\textbf {\bibinfo {volume} {55}},\ \bibinfo {pages} {443001} (\bibinfo {year} {2022})},\ \Eprint {http://arxiv.org/abs/2203.13011} {arXiv:2203.13011 [hep-th]} \BibitemShut {NoStop}%
	\bibitem [{\citenamefont {Yang}(2020)}]{Yang:2019vag}%
	  \BibitemOpen
	  \bibfield  {author} {\bibinfo {author} {\bibfnamefont {G.}~\bibnamefont {Yang}},\ }\href {\doibase 10.1007/s11433-019-1507-0} {\bibfield  {journal} {\bibinfo  {journal} {Sci. China Phys. Mech. Astron.}\ }\textbf {\bibinfo {volume} {63}},\ \bibinfo {pages} {270001} (\bibinfo {year} {2020})},\ \Eprint {http://arxiv.org/abs/1912.11454} {arXiv:1912.11454 [hep-th]} \BibitemShut {NoStop}%
	\bibitem [{\citenamefont {Alday}\ and\ \citenamefont {Maldacena}(2007)}]{Alday:2007he}%
	  \BibitemOpen
	  \bibfield  {author} {\bibinfo {author} {\bibfnamefont {L.~F.}\ \bibnamefont {Alday}}\ and\ \bibinfo {author} {\bibfnamefont {J.}~\bibnamefont {Maldacena}},\ }\href {\doibase 10.1088/1126-6708/2007/11/068} {\bibfield  {journal} {\bibinfo  {journal} {JHEP}\ }\textbf {\bibinfo {volume} {0711}},\ \bibinfo {pages} {068} (\bibinfo {year} {2007})},\ \Eprint {http://arxiv.org/abs/0710.1060} {arXiv:0710.1060 [hep-th]} \BibitemShut {NoStop}%
	\bibitem [{\citenamefont {Maldacena}\ and\ \citenamefont {Zhiboedov}(2010)}]{Maldacena:2010kp}%
	  \BibitemOpen
	  \bibfield  {author} {\bibinfo {author} {\bibfnamefont {J.}~\bibnamefont {Maldacena}}\ and\ \bibinfo {author} {\bibfnamefont {A.}~\bibnamefont {Zhiboedov}},\ }\href {\doibase 10.1007/jhep11(2010)104} {\bibfield  {journal} {\bibinfo  {journal} {JHEP}\ }\textbf {\bibinfo {volume} {11}},\ \bibinfo {pages} {104} (\bibinfo {year} {2010})},\ \Eprint {http://arxiv.org/abs/arXiv:1009.1139} {arXiv:1009.1139} \BibitemShut {NoStop}%
	\bibitem [{\citenamefont {Brandhuber}\ \emph {et~al.}(2011{\natexlab{a}})\citenamefont {Brandhuber}, \citenamefont {Spence}, \citenamefont {Travaglini},\ and\ \citenamefont {Yang}}]{Brandhuber:2010ad}%
	  \BibitemOpen
	  \bibfield  {author} {\bibinfo {author} {\bibfnamefont {A.}~\bibnamefont {Brandhuber}}, \bibinfo {author} {\bibfnamefont {B.}~\bibnamefont {Spence}}, \bibinfo {author} {\bibfnamefont {G.}~\bibnamefont {Travaglini}}, \ and\ \bibinfo {author} {\bibfnamefont {G.}~\bibnamefont {Yang}},\ }\href {\doibase https://doi.org/10.1007/JHEP01(2011)134} {\bibfield  {journal} {\bibinfo  {journal} {JHEP}\ }\textbf {\bibinfo {volume} {01}},\ \bibinfo {pages} {134} (\bibinfo {year} {2011}{\natexlab{a}})},\ \Eprint {http://arxiv.org/abs/arXiv:1011.1899} {arXiv:1011.1899} \BibitemShut {NoStop}%
	\bibitem [{\citenamefont {Brandhuber}\ \emph {et~al.}(2011{\natexlab{b}})\citenamefont {Brandhuber}, \citenamefont {Gurdogan}, \citenamefont {Mooney}, \citenamefont {Travaglini},\ and\ \citenamefont {Yang}}]{Brandhuber:2011tv}%
	  \BibitemOpen
	  \bibfield  {author} {\bibinfo {author} {\bibfnamefont {A.}~\bibnamefont {Brandhuber}}, \bibinfo {author} {\bibfnamefont {O.}~\bibnamefont {Gurdogan}}, \bibinfo {author} {\bibfnamefont {R.}~\bibnamefont {Mooney}}, \bibinfo {author} {\bibfnamefont {G.}~\bibnamefont {Travaglini}}, \ and\ \bibinfo {author} {\bibfnamefont {G.}~\bibnamefont {Yang}},\ }\href {\doibase 10.1007/JHEP10(2011)046} {\bibfield  {journal} {\bibinfo  {journal} {JHEP}\ }\textbf {\bibinfo {volume} {10}},\ \bibinfo {pages} {046} (\bibinfo {year} {2011}{\natexlab{b}})},\ \Eprint {http://arxiv.org/abs/1107.5067} {arXiv:1107.5067 [hep-th]} \BibitemShut {NoStop}%
	\bibitem [{\citenamefont {Ben-Israel}\ \emph {et~al.}(2018)\citenamefont {Ben-Israel}, \citenamefont {Tumanov},\ and\ \citenamefont {Sever}}]{Ben-Israel:2018ckc}%
	  \BibitemOpen
	  \bibfield  {author} {\bibinfo {author} {\bibfnamefont {R.}~\bibnamefont {Ben-Israel}}, \bibinfo {author} {\bibfnamefont {A.~G.}\ \bibnamefont {Tumanov}}, \ and\ \bibinfo {author} {\bibfnamefont {A.}~\bibnamefont {Sever}},\ }\href {\doibase 10.1007/JHEP08(2018)122} {\bibfield  {journal} {\bibinfo  {journal} {JHEP}\ }\textbf {\bibinfo {volume} {08}},\ \bibinfo {pages} {122} (\bibinfo {year} {2018})},\ \Eprint {http://arxiv.org/abs/1802.09395} {arXiv:1802.09395 [hep-th]} \BibitemShut {NoStop}%
	\bibitem [{\citenamefont {Bianchi}\ \emph {et~al.}(2019)\citenamefont {Bianchi}, \citenamefont {Brandhuber}, \citenamefont {Panerai},\ and\ \citenamefont {Travaglini}}]{Bianchi:2018rrj}%
	  \BibitemOpen
	  \bibfield  {author} {\bibinfo {author} {\bibfnamefont {L.}~\bibnamefont {Bianchi}}, \bibinfo {author} {\bibfnamefont {A.}~\bibnamefont {Brandhuber}}, \bibinfo {author} {\bibfnamefont {R.}~\bibnamefont {Panerai}}, \ and\ \bibinfo {author} {\bibfnamefont {G.}~\bibnamefont {Travaglini}},\ }\href {\doibase 10.1007/JHEP02(2019)134} {\bibfield  {journal} {\bibinfo  {journal} {JHEP}\ }\textbf {\bibinfo {volume} {02}},\ \bibinfo {pages} {134} (\bibinfo {year} {2019})},\ \Eprint {http://arxiv.org/abs/1812.10468} {arXiv:1812.10468 [hep-th]} \BibitemShut {NoStop}%
	\bibitem [{\citenamefont {Sever}\ \emph {et~al.}(2021{\natexlab{a}})\citenamefont {Sever}, \citenamefont {Tumanov},\ and\ \citenamefont {Wilhelm}}]{Sever:2020jjx}%
	  \BibitemOpen
	  \bibfield  {author} {\bibinfo {author} {\bibfnamefont {A.}~\bibnamefont {Sever}}, \bibinfo {author} {\bibfnamefont {A.~G.}\ \bibnamefont {Tumanov}}, \ and\ \bibinfo {author} {\bibfnamefont {M.}~\bibnamefont {Wilhelm}},\ }\href {\doibase 10.1103/PhysRevLett.126.031602} {\bibfield  {journal} {\bibinfo  {journal} {Phys. Rev. Lett.}\ }\textbf {\bibinfo {volume} {126}},\ \bibinfo {pages} {031602} (\bibinfo {year} {2021}{\natexlab{a}})},\ \Eprint {http://arxiv.org/abs/2009.11297} {arXiv:2009.11297 [hep-th]} \BibitemShut {NoStop}%
	\bibitem [{\citenamefont {Sever}\ \emph {et~al.}(2021{\natexlab{b}})\citenamefont {Sever}, \citenamefont {Tumanov},\ and\ \citenamefont {Wilhelm}}]{Sever:2021nsq}%
	  \BibitemOpen
	  \bibfield  {author} {\bibinfo {author} {\bibfnamefont {A.}~\bibnamefont {Sever}}, \bibinfo {author} {\bibfnamefont {A.~G.}\ \bibnamefont {Tumanov}}, \ and\ \bibinfo {author} {\bibfnamefont {M.}~\bibnamefont {Wilhelm}},\ }\href {\doibase 10.1007/JHEP10(2021)071} {\bibfield  {journal} {\bibinfo  {journal} {JHEP}\ }\textbf {\bibinfo {volume} {10}},\ \bibinfo {pages} {071} (\bibinfo {year} {2021}{\natexlab{b}})},\ \Eprint {http://arxiv.org/abs/2105.13367} {arXiv:2105.13367 [hep-th]} \BibitemShut {NoStop}%
	\bibitem [{\citenamefont {Sever}\ \emph {et~al.}(2022)\citenamefont {Sever}, \citenamefont {Tumanov},\ and\ \citenamefont {Wilhelm}}]{Sever:2021xga}%
	  \BibitemOpen
	  \bibfield  {author} {\bibinfo {author} {\bibfnamefont {A.}~\bibnamefont {Sever}}, \bibinfo {author} {\bibfnamefont {A.~G.}\ \bibnamefont {Tumanov}}, \ and\ \bibinfo {author} {\bibfnamefont {M.}~\bibnamefont {Wilhelm}},\ }\href {\doibase 10.1007/JHEP03(2022)128} {\bibfield  {journal} {\bibinfo  {journal} {JHEP}\ }\textbf {\bibinfo {volume} {03}},\ \bibinfo {pages} {128} (\bibinfo {year} {2022})},\ \Eprint {http://arxiv.org/abs/2112.10569} {arXiv:2112.10569 [hep-th]} \BibitemShut {NoStop}%
	\bibitem [{\citenamefont {Alday}\ \emph {et~al.}(2011)\citenamefont {Alday}, \citenamefont {Gaiotto}, \citenamefont {Maldacena}, \citenamefont {Sever},\ and\ \citenamefont {Vieira}}]{Alday:2010ku}%
	  \BibitemOpen
	  \bibfield  {author} {\bibinfo {author} {\bibfnamefont {L.~F.}\ \bibnamefont {Alday}}, \bibinfo {author} {\bibfnamefont {D.}~\bibnamefont {Gaiotto}}, \bibinfo {author} {\bibfnamefont {J.}~\bibnamefont {Maldacena}}, \bibinfo {author} {\bibfnamefont {A.}~\bibnamefont {Sever}}, \ and\ \bibinfo {author} {\bibfnamefont {P.}~\bibnamefont {Vieira}},\ }\href {\doibase 10.1007/JHEP04(2011)088} {\bibfield  {journal} {\bibinfo  {journal} {JHEP}\ }\textbf {\bibinfo {volume} {04}},\ \bibinfo {pages} {088} (\bibinfo {year} {2011})},\ \Eprint {http://arxiv.org/abs/1006.2788} {arXiv:1006.2788 [hep-th]} \BibitemShut {NoStop}%
	\bibitem [{\citenamefont {Basso}\ \emph {et~al.}(2013)\citenamefont {Basso}, \citenamefont {Sever},\ and\ \citenamefont {Vieira}}]{Basso:2013vsa}%
	  \BibitemOpen
	  \bibfield  {author} {\bibinfo {author} {\bibfnamefont {B.}~\bibnamefont {Basso}}, \bibinfo {author} {\bibfnamefont {A.}~\bibnamefont {Sever}}, \ and\ \bibinfo {author} {\bibfnamefont {P.}~\bibnamefont {Vieira}},\ }\href {\doibase 10.1103/PhysRevLett.111.091602} {\bibfield  {journal} {\bibinfo  {journal} {Phys. Rev. Lett.}\ }\textbf {\bibinfo {volume} {111}},\ \bibinfo {pages} {091602} (\bibinfo {year} {2013})},\ \Eprint {http://arxiv.org/abs/1303.1396} {arXiv:1303.1396 [hep-th]} \BibitemShut {NoStop}%
	\bibitem [{\citenamefont {Basso}\ \emph {et~al.}(2014{\natexlab{a}})\citenamefont {Basso}, \citenamefont {Sever},\ and\ \citenamefont {Vieira}}]{Basso:2013aha}%
	  \BibitemOpen
	  \bibfield  {author} {\bibinfo {author} {\bibfnamefont {B.}~\bibnamefont {Basso}}, \bibinfo {author} {\bibfnamefont {A.}~\bibnamefont {Sever}}, \ and\ \bibinfo {author} {\bibfnamefont {P.}~\bibnamefont {Vieira}},\ }\href {\doibase 10.1007/JHEP01(2014)008} {\bibfield  {journal} {\bibinfo  {journal} {JHEP}\ }\textbf {\bibinfo {volume} {01}},\ \bibinfo {pages} {008} (\bibinfo {year} {2014}{\natexlab{a}})},\ \Eprint {http://arxiv.org/abs/1306.2058} {arXiv:1306.2058 [hep-th]} \BibitemShut {NoStop}%
	\bibitem [{\citenamefont {Basso}\ \emph {et~al.}(2014{\natexlab{b}})\citenamefont {Basso}, \citenamefont {Sever},\ and\ \citenamefont {Vieira}}]{Basso:2014koa}%
	  \BibitemOpen
	  \bibfield  {author} {\bibinfo {author} {\bibfnamefont {B.}~\bibnamefont {Basso}}, \bibinfo {author} {\bibfnamefont {A.}~\bibnamefont {Sever}}, \ and\ \bibinfo {author} {\bibfnamefont {P.}~\bibnamefont {Vieira}},\ }\href {\doibase 10.1007/JHEP08(2014)085} {\bibfield  {journal} {\bibinfo  {journal} {JHEP}\ }\textbf {\bibinfo {volume} {08}},\ \bibinfo {pages} {085} (\bibinfo {year} {2014}{\natexlab{b}})},\ \Eprint {http://arxiv.org/abs/1402.3307} {arXiv:1402.3307 [hep-th]} \BibitemShut {NoStop}%
	\bibitem [{\citenamefont {Basso}\ \emph {et~al.}(2014{\natexlab{c}})\citenamefont {Basso}, \citenamefont {Sever},\ and\ \citenamefont {Vieira}}]{Basso:2014nra}%
	  \BibitemOpen
	  \bibfield  {author} {\bibinfo {author} {\bibfnamefont {B.}~\bibnamefont {Basso}}, \bibinfo {author} {\bibfnamefont {A.}~\bibnamefont {Sever}}, \ and\ \bibinfo {author} {\bibfnamefont {P.}~\bibnamefont {Vieira}},\ }\href {\doibase 10.1007/JHEP09(2014)149} {\bibfield  {journal} {\bibinfo  {journal} {JHEP}\ }\textbf {\bibinfo {volume} {09}},\ \bibinfo {pages} {149} (\bibinfo {year} {2014}{\natexlab{c}})},\ \Eprint {http://arxiv.org/abs/1407.1736} {arXiv:1407.1736 [hep-th]} \BibitemShut {NoStop}%
	\bibitem [{\citenamefont {Basso}\ \emph {et~al.}(2015{\natexlab{a}})\citenamefont {Basso}, \citenamefont {Caetano}, \citenamefont {Cordova}, \citenamefont {Sever},\ and\ \citenamefont {Vieira}}]{Basso:2014hfa}%
	  \BibitemOpen
	  \bibfield  {author} {\bibinfo {author} {\bibfnamefont {B.}~\bibnamefont {Basso}}, \bibinfo {author} {\bibfnamefont {J.}~\bibnamefont {Caetano}}, \bibinfo {author} {\bibfnamefont {L.}~\bibnamefont {Cordova}}, \bibinfo {author} {\bibfnamefont {A.}~\bibnamefont {Sever}}, \ and\ \bibinfo {author} {\bibfnamefont {P.}~\bibnamefont {Vieira}},\ }\href {\doibase 10.1007/JHEP08(2015)018} {\bibfield  {journal} {\bibinfo  {journal} {JHEP}\ }\textbf {\bibinfo {volume} {08}},\ \bibinfo {pages} {018} (\bibinfo {year} {2015}{\natexlab{a}})},\ \Eprint {http://arxiv.org/abs/1412.1132} {arXiv:1412.1132 [hep-th]} \BibitemShut {NoStop}%
	\bibitem [{\citenamefont {Basso}\ \emph {et~al.}(2015{\natexlab{b}})\citenamefont {Basso}, \citenamefont {Caetano}, \citenamefont {Cordova}, \citenamefont {Sever},\ and\ \citenamefont {Vieira}}]{Basso:2015rta}%
	  \BibitemOpen
	  \bibfield  {author} {\bibinfo {author} {\bibfnamefont {B.}~\bibnamefont {Basso}}, \bibinfo {author} {\bibfnamefont {J.}~\bibnamefont {Caetano}}, \bibinfo {author} {\bibfnamefont {L.}~\bibnamefont {Cordova}}, \bibinfo {author} {\bibfnamefont {A.}~\bibnamefont {Sever}}, \ and\ \bibinfo {author} {\bibfnamefont {P.}~\bibnamefont {Vieira}},\ }\href {\doibase 10.1007/JHEP12(2015)088} {\bibfield  {journal} {\bibinfo  {journal} {JHEP}\ }\textbf {\bibinfo {volume} {12}},\ \bibinfo {pages} {088} (\bibinfo {year} {2015}{\natexlab{b}})},\ \Eprint {http://arxiv.org/abs/1508.02987} {arXiv:1508.02987 [hep-th]} \BibitemShut {NoStop}%
	\bibitem [{\citenamefont {Basso}\ \emph {et~al.}(2016)\citenamefont {Basso}, \citenamefont {Sever},\ and\ \citenamefont {Vieira}}]{Basso:2015uxa}%
	  \BibitemOpen
	  \bibfield  {author} {\bibinfo {author} {\bibfnamefont {B.}~\bibnamefont {Basso}}, \bibinfo {author} {\bibfnamefont {A.}~\bibnamefont {Sever}}, \ and\ \bibinfo {author} {\bibfnamefont {P.}~\bibnamefont {Vieira}},\ }\href {\doibase 10.1088/1751-8113/49/41/41LT01} {\bibfield  {journal} {\bibinfo  {journal} {J. Phys. A}\ }\textbf {\bibinfo {volume} {49}},\ \bibinfo {pages} {41LT01} (\bibinfo {year} {2016})},\ \Eprint {http://arxiv.org/abs/1508.03045} {arXiv:1508.03045 [hep-th]} \BibitemShut {NoStop}%
	\bibitem [{\citenamefont {Belitsky}(2015{\natexlab{a}})}]{Belitsky:2014sla}%
	  \BibitemOpen
	  \bibfield  {author} {\bibinfo {author} {\bibfnamefont {A.}~\bibnamefont {Belitsky}},\ }\href {\doibase 10.1016/j.nuclphysb.2015.05.002} {\bibfield  {journal} {\bibinfo  {journal} {Nucl. Phys. B}\ }\textbf {\bibinfo {volume} {896}},\ \bibinfo {pages} {493} (\bibinfo {year} {2015}{\natexlab{a}})},\ \Eprint {http://arxiv.org/abs/1407.2853} {arXiv:1407.2853 [hep-th]} \BibitemShut {NoStop}%
	\bibitem [{\citenamefont {Belitsky}(2015{\natexlab{b}})}]{Belitsky:2014lta}%
	  \BibitemOpen
	  \bibfield  {author} {\bibinfo {author} {\bibfnamefont {A.}~\bibnamefont {Belitsky}},\ }\href {\doibase 10.1016/j.nuclphysb.2015.02.025} {\bibfield  {journal} {\bibinfo  {journal} {Nucl. Phys. B}\ }\textbf {\bibinfo {volume} {894}},\ \bibinfo {pages} {108} (\bibinfo {year} {2015}{\natexlab{b}})},\ \Eprint {http://arxiv.org/abs/1410.2534} {arXiv:1410.2534 [hep-th]} \BibitemShut {NoStop}%
	\bibitem [{\citenamefont {Belitsky}(2017)}]{Belitsky:2016vyq}%
	  \BibitemOpen
	  \bibfield  {author} {\bibinfo {author} {\bibfnamefont {A.}~\bibnamefont {Belitsky}},\ }\href {\doibase 10.1016/j.nuclphysb.2017.08.011} {\bibfield  {journal} {\bibinfo  {journal} {Nucl. Phys. B}\ }\textbf {\bibinfo {volume} {923}},\ \bibinfo {pages} {588} (\bibinfo {year} {2017})},\ \Eprint {http://arxiv.org/abs/1607.06555} {arXiv:1607.06555 [hep-th]} \BibitemShut {NoStop}%
	\bibitem [{\citenamefont {Dixon}\ \emph {et~al.}(2021)\citenamefont {Dixon}, \citenamefont {McLeod},\ and\ \citenamefont {Wilhelm}}]{Dixon:2020bbt}%
	  \BibitemOpen
	  \bibfield  {author} {\bibinfo {author} {\bibfnamefont {L.~J.}\ \bibnamefont {Dixon}}, \bibinfo {author} {\bibfnamefont {A.~J.}\ \bibnamefont {McLeod}}, \ and\ \bibinfo {author} {\bibfnamefont {M.}~\bibnamefont {Wilhelm}},\ }\href {\doibase 10.1007/JHEP04(2021)147} {\bibfield  {journal} {\bibinfo  {journal} {JHEP}\ }\textbf {\bibinfo {volume} {04}},\ \bibinfo {pages} {147} (\bibinfo {year} {2021})},\ \Eprint {http://arxiv.org/abs/2012.12286} {arXiv:2012.12286 [hep-th]} \BibitemShut {NoStop}%
	\bibitem [{\citenamefont {Dixon}\ \emph {et~al.}(2022{\natexlab{a}})\citenamefont {Dixon}, \citenamefont {Gurdogan}, \citenamefont {McLeod},\ and\ \citenamefont {Wilhelm}}]{Dixon:2022rse}%
	  \BibitemOpen
	  \bibfield  {author} {\bibinfo {author} {\bibfnamefont {L.~J.}\ \bibnamefont {Dixon}}, \bibinfo {author} {\bibfnamefont {O.}~\bibnamefont {Gurdogan}}, \bibinfo {author} {\bibfnamefont {A.~J.}\ \bibnamefont {McLeod}}, \ and\ \bibinfo {author} {\bibfnamefont {M.}~\bibnamefont {Wilhelm}},\ }\href {\doibase 10.1007/JHEP07(2022)153} {\bibfield  {journal} {\bibinfo  {journal} {JHEP}\ }\textbf {\bibinfo {volume} {07}},\ \bibinfo {pages} {153} (\bibinfo {year} {2022}{\natexlab{a}})},\ \Eprint {http://arxiv.org/abs/2204.11901} {arXiv:2204.11901 [hep-th]} \BibitemShut {NoStop}%
	\bibitem [{\citenamefont {Dixon}\ \emph {et~al.}(2011)\citenamefont {Dixon}, \citenamefont {Drummond},\ and\ \citenamefont {Henn}}]{Dixon:2011pw}%
	  \BibitemOpen
	  \bibfield  {author} {\bibinfo {author} {\bibfnamefont {L.~J.}\ \bibnamefont {Dixon}}, \bibinfo {author} {\bibfnamefont {J.~M.}\ \bibnamefont {Drummond}}, \ and\ \bibinfo {author} {\bibfnamefont {J.~M.}\ \bibnamefont {Henn}},\ }\href {\doibase 10.1007/JHEP11(2011)023} {\bibfield  {journal} {\bibinfo  {journal} {JHEP}\ }\textbf {\bibinfo {volume} {11}},\ \bibinfo {pages} {023} (\bibinfo {year} {2011})},\ \Eprint {http://arxiv.org/abs/1108.4461} {arXiv:1108.4461 [hep-th]} \BibitemShut {NoStop}%
	\bibitem [{\citenamefont {Brandhuber}\ \emph {et~al.}(2012)\citenamefont {Brandhuber}, \citenamefont {Travaglini},\ and\ \citenamefont {Yang}}]{Brandhuber:2012vm}%
	  \BibitemOpen
	  \bibfield  {author} {\bibinfo {author} {\bibfnamefont {A.}~\bibnamefont {Brandhuber}}, \bibinfo {author} {\bibfnamefont {G.}~\bibnamefont {Travaglini}}, \ and\ \bibinfo {author} {\bibfnamefont {G.}~\bibnamefont {Yang}},\ }\href {\doibase 10.1007/JHEP05(2012)082} {\bibfield  {journal} {\bibinfo  {journal} {JHEP}\ }\textbf {\bibinfo {volume} {05}},\ \bibinfo {pages} {082} (\bibinfo {year} {2012})},\ \Eprint {http://arxiv.org/abs/1201.4170} {arXiv:1201.4170 [hep-th]} \BibitemShut {NoStop}%
	\bibitem [{\citenamefont {Dixon}\ \emph {et~al.}(2013)\citenamefont {Dixon}, \citenamefont {Drummond}, \citenamefont {von Hippel},\ and\ \citenamefont {Pennington}}]{Dixon:2013eka}%
	  \BibitemOpen
	  \bibfield  {author} {\bibinfo {author} {\bibfnamefont {L.~J.}\ \bibnamefont {Dixon}}, \bibinfo {author} {\bibfnamefont {J.~M.}\ \bibnamefont {Drummond}}, \bibinfo {author} {\bibfnamefont {M.}~\bibnamefont {von Hippel}}, \ and\ \bibinfo {author} {\bibfnamefont {J.}~\bibnamefont {Pennington}},\ }\href {\doibase 10.1007/JHEP12(2013)049} {\bibfield  {journal} {\bibinfo  {journal} {JHEP}\ }\textbf {\bibinfo {volume} {12}},\ \bibinfo {pages} {049} (\bibinfo {year} {2013})},\ \Eprint {http://arxiv.org/abs/1308.2276} {arXiv:1308.2276 [hep-th]} \BibitemShut {NoStop}%
	\bibitem [{\citenamefont {Dixon}\ and\ \citenamefont {von Hippel}(2014)}]{Dixon:2014iba}%
	  \BibitemOpen
	  \bibfield  {author} {\bibinfo {author} {\bibfnamefont {L.~J.}\ \bibnamefont {Dixon}}\ and\ \bibinfo {author} {\bibfnamefont {M.}~\bibnamefont {von Hippel}},\ }\href {\doibase 10.1007/JHEP10(2014)065} {\bibfield  {journal} {\bibinfo  {journal} {JHEP}\ }\textbf {\bibinfo {volume} {10}},\ \bibinfo {pages} {065} (\bibinfo {year} {2014})},\ \Eprint {http://arxiv.org/abs/1408.1505} {arXiv:1408.1505 [hep-th]} \BibitemShut {NoStop}%
	\bibitem [{\citenamefont {Golden}\ and\ \citenamefont {Spradlin}(2015)}]{Golden:2014pua}%
	  \BibitemOpen
	  \bibfield  {author} {\bibinfo {author} {\bibfnamefont {J.}~\bibnamefont {Golden}}\ and\ \bibinfo {author} {\bibfnamefont {M.}~\bibnamefont {Spradlin}},\ }\href {\doibase 10.1007/JHEP02(2015)002} {\bibfield  {journal} {\bibinfo  {journal} {JHEP}\ }\textbf {\bibinfo {volume} {02}},\ \bibinfo {pages} {002} (\bibinfo {year} {2015})},\ \Eprint {http://arxiv.org/abs/1411.3289} {arXiv:1411.3289 [hep-th]} \BibitemShut {NoStop}%
	\bibitem [{\citenamefont {Drummond}\ \emph {et~al.}(2015)\citenamefont {Drummond}, \citenamefont {Papathanasiou},\ and\ \citenamefont {Spradlin}}]{Drummond:2014ffa}%
	  \BibitemOpen
	  \bibfield  {author} {\bibinfo {author} {\bibfnamefont {J.~M.}\ \bibnamefont {Drummond}}, \bibinfo {author} {\bibfnamefont {G.}~\bibnamefont {Papathanasiou}}, \ and\ \bibinfo {author} {\bibfnamefont {M.}~\bibnamefont {Spradlin}},\ }\href {\doibase 10.1007/JHEP03(2015)072} {\bibfield  {journal} {\bibinfo  {journal} {JHEP}\ }\textbf {\bibinfo {volume} {03}},\ \bibinfo {pages} {072} (\bibinfo {year} {2015})},\ \Eprint {http://arxiv.org/abs/1412.3763} {arXiv:1412.3763 [hep-th]} \BibitemShut {NoStop}%
	\bibitem [{\citenamefont {Caron-Huot}\ \emph {et~al.}(2016)\citenamefont {Caron-Huot}, \citenamefont {Dixon}, \citenamefont {McLeod},\ and\ \citenamefont {von Hippel}}]{Caron-Huot:2016owq}%
	  \BibitemOpen
	  \bibfield  {author} {\bibinfo {author} {\bibfnamefont {S.}~\bibnamefont {Caron-Huot}}, \bibinfo {author} {\bibfnamefont {L.~J.}\ \bibnamefont {Dixon}}, \bibinfo {author} {\bibfnamefont {A.}~\bibnamefont {McLeod}}, \ and\ \bibinfo {author} {\bibfnamefont {M.}~\bibnamefont {von Hippel}},\ }\href {\doibase 10.1103/PhysRevLett.117.241601} {\bibfield  {journal} {\bibinfo  {journal} {Phys. Rev. Lett.}\ }\textbf {\bibinfo {volume} {117}},\ \bibinfo {pages} {241601} (\bibinfo {year} {2016})},\ \Eprint {http://arxiv.org/abs/1609.00669} {arXiv:1609.00669 [hep-th]} \BibitemShut {NoStop}%
	\bibitem [{\citenamefont {Dixon}\ \emph {et~al.}(2017)\citenamefont {Dixon}, \citenamefont {Drummond}, \citenamefont {Harrington}, \citenamefont {McLeod}, \citenamefont {Papathanasiou},\ and\ \citenamefont {Spradlin}}]{Dixon:2016nkn}%
	  \BibitemOpen
	  \bibfield  {author} {\bibinfo {author} {\bibfnamefont {L.~J.}\ \bibnamefont {Dixon}}, \bibinfo {author} {\bibfnamefont {J.}~\bibnamefont {Drummond}}, \bibinfo {author} {\bibfnamefont {T.}~\bibnamefont {Harrington}}, \bibinfo {author} {\bibfnamefont {A.~J.}\ \bibnamefont {McLeod}}, \bibinfo {author} {\bibfnamefont {G.}~\bibnamefont {Papathanasiou}}, \ and\ \bibinfo {author} {\bibfnamefont {M.}~\bibnamefont {Spradlin}},\ }\href {\doibase 10.1007/JHEP02(2017)137} {\bibfield  {journal} {\bibinfo  {journal} {JHEP}\ }\textbf {\bibinfo {volume} {02}},\ \bibinfo {pages} {137} (\bibinfo {year} {2017})},\ \Eprint {http://arxiv.org/abs/1612.08976} {arXiv:1612.08976 [hep-th]} \BibitemShut {NoStop}%
	\bibitem [{\citenamefont {Drummond}\ \emph {et~al.}(2019)\citenamefont {Drummond}, \citenamefont {Foster}, \citenamefont {G\"urdo\u{g}an},\ and\ \citenamefont {Papathanasiou}}]{Drummond:2018caf}%
	  \BibitemOpen
	  \bibfield  {author} {\bibinfo {author} {\bibfnamefont {J.}~\bibnamefont {Drummond}}, \bibinfo {author} {\bibfnamefont {J.}~\bibnamefont {Foster}}, \bibinfo {author} {\bibfnamefont {O.}~\bibnamefont {G\"urdo\u{g}an}}, \ and\ \bibinfo {author} {\bibfnamefont {G.}~\bibnamefont {Papathanasiou}},\ }\href {\doibase 10.1007/JHEP03(2019)087} {\bibfield  {journal} {\bibinfo  {journal} {JHEP}\ }\textbf {\bibinfo {volume} {03}},\ \bibinfo {pages} {087} (\bibinfo {year} {2019})},\ \Eprint {http://arxiv.org/abs/1812.04640} {arXiv:1812.04640 [hep-th]} \BibitemShut {NoStop}%
	\bibitem [{\citenamefont {Caron-Huot}\ \emph {et~al.}(2019)\citenamefont {Caron-Huot}, \citenamefont {Dixon}, \citenamefont {Dulat}, \citenamefont {von Hippel}, \citenamefont {McLeod},\ and\ \citenamefont {Papathanasiou}}]{Caron-Huot:2019vjl}%
	  \BibitemOpen
	  \bibfield  {author} {\bibinfo {author} {\bibfnamefont {S.}~\bibnamefont {Caron-Huot}}, \bibinfo {author} {\bibfnamefont {L.~J.}\ \bibnamefont {Dixon}}, \bibinfo {author} {\bibfnamefont {F.}~\bibnamefont {Dulat}}, \bibinfo {author} {\bibfnamefont {M.}~\bibnamefont {von Hippel}}, \bibinfo {author} {\bibfnamefont {A.~J.}\ \bibnamefont {McLeod}}, \ and\ \bibinfo {author} {\bibfnamefont {G.}~\bibnamefont {Papathanasiou}},\ }\href {\doibase 10.1007/JHEP08(2019)016} {\bibfield  {journal} {\bibinfo  {journal} {JHEP}\ }\textbf {\bibinfo {volume} {08}},\ \bibinfo {pages} {016} (\bibinfo {year} {2019})},\ \Eprint {http://arxiv.org/abs/1903.10890} {arXiv:1903.10890 [hep-th]} \BibitemShut {NoStop}%
	\bibitem [{\citenamefont {Dixon}\ and\ \citenamefont {Liu}(2020)}]{Dixon:2020cnr}%
	  \BibitemOpen
	  \bibfield  {author} {\bibinfo {author} {\bibfnamefont {L.~J.}\ \bibnamefont {Dixon}}\ and\ \bibinfo {author} {\bibfnamefont {Y.-T.}\ \bibnamefont {Liu}},\ }\href {\doibase 10.1007/JHEP10(2020)031} {\bibfield  {journal} {\bibinfo  {journal} {JHEP}\ }\textbf {\bibinfo {volume} {10}},\ \bibinfo {pages} {031} (\bibinfo {year} {2020})},\ \Eprint {http://arxiv.org/abs/2007.12966} {arXiv:2007.12966 [hep-th]} \BibitemShut {NoStop}%
	\bibitem [{\citenamefont {He}\ \emph {et~al.}(2020)\citenamefont {He}, \citenamefont {Li},\ and\ \citenamefont {Zhang}}]{Zhang:2019vnm}%
	  \BibitemOpen
	  \bibfield  {author} {\bibinfo {author} {\bibfnamefont {S.}~\bibnamefont {He}}, \bibinfo {author} {\bibfnamefont {Z.}~\bibnamefont {Li}}, \ and\ \bibinfo {author} {\bibfnamefont {C.}~\bibnamefont {Zhang}},\ }\href {\doibase 10.1103/PhysRevD.101.061701} {\bibfield  {journal} {\bibinfo  {journal} {Phys. Rev. D}\ }\textbf {\bibinfo {volume} {101}},\ \bibinfo {pages} {061701} (\bibinfo {year} {2020})},\ \Eprint {http://arxiv.org/abs/1911.01290} {arXiv:1911.01290 [hep-th]} \BibitemShut {NoStop}%
	\bibitem [{\citenamefont {He}\ \emph {et~al.}(2021)\citenamefont {He}, \citenamefont {Li},\ and\ \citenamefont {Zhang}}]{He:2020vob}%
	  \BibitemOpen
	  \bibfield  {author} {\bibinfo {author} {\bibfnamefont {S.}~\bibnamefont {He}}, \bibinfo {author} {\bibfnamefont {Z.}~\bibnamefont {Li}}, \ and\ \bibinfo {author} {\bibfnamefont {C.}~\bibnamefont {Zhang}},\ }\href {\doibase 10.1007/JHEP03(2021)278} {\bibfield  {journal} {\bibinfo  {journal} {JHEP}\ }\textbf {\bibinfo {volume} {03}},\ \bibinfo {pages} {278} (\bibinfo {year} {2021})},\ \Eprint {http://arxiv.org/abs/2009.11471} {arXiv:2009.11471 [hep-th]} \BibitemShut {NoStop}%
	\bibitem [{\citenamefont {Golden}\ and\ \citenamefont {Mcleod}(2021)}]{Golden:2021ggj}%
	  \BibitemOpen
	  \bibfield  {author} {\bibinfo {author} {\bibfnamefont {J.}~\bibnamefont {Golden}}\ and\ \bibinfo {author} {\bibfnamefont {A.~J.}\ \bibnamefont {Mcleod}},\ }\href@noop {} {\  (\bibinfo {year} {2021})},\ \Eprint {http://arxiv.org/abs/2104.14194} {arXiv:2104.14194 [hep-th]} \BibitemShut {NoStop}%
	\bibitem [{\citenamefont {Dixon}\ \emph {et~al.}(2022{\natexlab{b}})\citenamefont {Dixon}, \citenamefont {Gurdogan}, \citenamefont {McLeod},\ and\ \citenamefont {Wilhelm}}]{Dixon:2021tdw}%
	  \BibitemOpen
	  \bibfield  {author} {\bibinfo {author} {\bibfnamefont {L.~J.}\ \bibnamefont {Dixon}}, \bibinfo {author} {\bibfnamefont {O.}~\bibnamefont {Gurdogan}}, \bibinfo {author} {\bibfnamefont {A.~J.}\ \bibnamefont {McLeod}}, \ and\ \bibinfo {author} {\bibfnamefont {M.}~\bibnamefont {Wilhelm}},\ }\href {\doibase 10.1103/PhysRevLett.128.111602} {\bibfield  {journal} {\bibinfo  {journal} {Phys. Rev. Lett.}\ }\textbf {\bibinfo {volume} {128}},\ \bibinfo {pages} {111602} (\bibinfo {year} {2022}{\natexlab{b}})},\ \Eprint {http://arxiv.org/abs/2112.06243} {arXiv:2112.06243 [hep-th]} \BibitemShut {NoStop}%
	\bibitem [{\citenamefont {Lin}\ \emph {et~al.}(2021)\citenamefont {Lin}, \citenamefont {Yang},\ and\ \citenamefont {Zhang}}]{Lin:2021lqo}%
	  \BibitemOpen
	  \bibfield  {author} {\bibinfo {author} {\bibfnamefont {G.}~\bibnamefont {Lin}}, \bibinfo {author} {\bibfnamefont {G.}~\bibnamefont {Yang}}, \ and\ \bibinfo {author} {\bibfnamefont {S.}~\bibnamefont {Zhang}},\ }\href@noop {} {\  (\bibinfo {year} {2021})},\ \Eprint {http://arxiv.org/abs/2112.09123} {arXiv:2112.09123 [hep-th]} \BibitemShut {NoStop}%
	\bibitem [{\citenamefont {Drummond}\ \emph {et~al.}(2010)\citenamefont {Drummond}, \citenamefont {Henn}, \citenamefont {Korchemsky},\ and\ \citenamefont {Sokatchev}}]{Drummond:2007au}%
	  \BibitemOpen
	  \bibfield  {author} {\bibinfo {author} {\bibfnamefont {J.~M.}\ \bibnamefont {Drummond}}, \bibinfo {author} {\bibfnamefont {J.}~\bibnamefont {Henn}}, \bibinfo {author} {\bibfnamefont {G.~P.}\ \bibnamefont {Korchemsky}}, \ and\ \bibinfo {author} {\bibfnamefont {E.}~\bibnamefont {Sokatchev}},\ }\href {\doibase 10.1016/j.nuclphysb.2009.10.013} {\bibfield  {journal} {\bibinfo  {journal} {Nucl. Phys.}\ }\textbf {\bibinfo {volume} {B826}},\ \bibinfo {pages} {337} (\bibinfo {year} {2010})},\ \Eprint {http://arxiv.org/abs/0712.1223} {arXiv:0712.1223 [hep-th]} \BibitemShut {NoStop}%
	\bibitem [{\citenamefont {Guo}\ \emph {et~al.}(2021)\citenamefont {Guo}, \citenamefont {Wang},\ and\ \citenamefont {Yang}}]{Guo:2021bym}%
	  \BibitemOpen
	  \bibfield  {author} {\bibinfo {author} {\bibfnamefont {Y.}~\bibnamefont {Guo}}, \bibinfo {author} {\bibfnamefont {L.}~\bibnamefont {Wang}}, \ and\ \bibinfo {author} {\bibfnamefont {G.}~\bibnamefont {Yang}},\ }\href {\doibase 10.1103/PhysRevLett.127.151602} {\bibfield  {journal} {\bibinfo  {journal} {Phys. Rev. Lett.}\ }\textbf {\bibinfo {volume} {127}},\ \bibinfo {pages} {151602} (\bibinfo {year} {2021})},\ \Eprint {http://arxiv.org/abs/2106.01374} {arXiv:2106.01374 [hep-th]} \BibitemShut {NoStop}%
	\bibitem [{\citenamefont {Guo}\ \emph {et~al.}(2022)\citenamefont {Guo}, \citenamefont {Jin}, \citenamefont {Wang},\ and\ \citenamefont {Yang}}]{Guo:2022pdw}%
	  \BibitemOpen
	  \bibfield  {author} {\bibinfo {author} {\bibfnamefont {Y.}~\bibnamefont {Guo}}, \bibinfo {author} {\bibfnamefont {Q.}~\bibnamefont {Jin}}, \bibinfo {author} {\bibfnamefont {L.}~\bibnamefont {Wang}}, \ and\ \bibinfo {author} {\bibfnamefont {G.}~\bibnamefont {Yang}},\ }\href@noop {} {\  (\bibinfo {year} {2022})},\ \Eprint {http://arxiv.org/abs/2205.12969} {arXiv:2205.12969 [hep-th]} \BibitemShut {NoStop}%
	\bibitem [{\citenamefont {BERN}\ \emph {et~al.}(1994)\citenamefont {BERN}, \citenamefont {DIXON}, \citenamefont {DUNBAR},\ and\ \citenamefont {KOSOWER}}]{Bern:1994zx}%
	  \BibitemOpen
	  \bibfield  {author} {\bibinfo {author} {\bibfnamefont {Z.}~\bibnamefont {BERN}}, \bibinfo {author} {\bibfnamefont {L.}~\bibnamefont {DIXON}}, \bibinfo {author} {\bibfnamefont {D.~C.}\ \bibnamefont {DUNBAR}}, \ and\ \bibinfo {author} {\bibfnamefont {D.~A.}\ \bibnamefont {KOSOWER}},\ }\href {\doibase 10.1016/0550-3213(94)90179-1} {\bibfield  {journal} {\bibinfo  {journal} {Nuclear Physics}\ }\textbf {\bibinfo {volume} {B425}},\ \bibinfo {pages} {217} (\bibinfo {year} {1994})},\ \Eprint {http://arxiv.org/abs/arXiv:hep-ph/9403226} {arXiv:hep-ph/9403226} \BibitemShut {NoStop}%
	\bibitem [{\citenamefont {Bern}\ \emph {et~al.}(1995)\citenamefont {Bern}, \citenamefont {Dixon}, \citenamefont {Dunbar},\ and\ \citenamefont {Kosower}}]{Bern:1994cg}%
	  \BibitemOpen
	  \bibfield  {author} {\bibinfo {author} {\bibfnamefont {Z.}~\bibnamefont {Bern}}, \bibinfo {author} {\bibfnamefont {L.}~\bibnamefont {Dixon}}, \bibinfo {author} {\bibfnamefont {D.}~\bibnamefont {Dunbar}}, \ and\ \bibinfo {author} {\bibfnamefont {D.}~\bibnamefont {Kosower}},\ }\href {\doibase 10.1016/0550-3213(94)00488-Z} {\bibfield  {journal} {\bibinfo  {journal} {Nuclear Physics B}\ }\textbf {\bibinfo {volume} {435}},\ \bibinfo {pages} {59} (\bibinfo {year} {1995})},\ \Eprint {http://arxiv.org/abs/arXiv:hep-ph/9409265} {arXiv:hep-ph/9409265} \BibitemShut {NoStop}%
	\bibitem [{\citenamefont {Britto}\ \emph {et~al.}(2005)\citenamefont {Britto}, \citenamefont {Cachazo},\ and\ \citenamefont {Feng}}]{Britto:2004nc}%
	  \BibitemOpen
	  \bibfield  {author} {\bibinfo {author} {\bibfnamefont {R.}~\bibnamefont {Britto}}, \bibinfo {author} {\bibfnamefont {F.}~\bibnamefont {Cachazo}}, \ and\ \bibinfo {author} {\bibfnamefont {B.}~\bibnamefont {Feng}},\ }\href {\doibase 10.1016/j.nuclphysb.2005.07.014} {\bibfield  {journal} {\bibinfo  {journal} {Nucl.Phys.}\ }\textbf {\bibinfo {volume} {B725}},\ \bibinfo {pages} {275} (\bibinfo {year} {2005})},\ \Eprint {http://arxiv.org/abs/hep-th/0412103} {arXiv:hep-th/0412103 [hep-th]} \BibitemShut {NoStop}%
	\bibitem [{\citenamefont {Papadopoulos}\ \emph {et~al.}(2016)\citenamefont {Papadopoulos}, \citenamefont {Tommasini},\ and\ \citenamefont {Wever}}]{Papadopoulos:2015jft}%
	  \BibitemOpen
	  \bibfield  {author} {\bibinfo {author} {\bibfnamefont {C.~G.}\ \bibnamefont {Papadopoulos}}, \bibinfo {author} {\bibfnamefont {D.}~\bibnamefont {Tommasini}}, \ and\ \bibinfo {author} {\bibfnamefont {C.}~\bibnamefont {Wever}},\ }\href {\doibase 10.1007/JHEP04(2016)078} {\bibfield  {journal} {\bibinfo  {journal} {JHEP}\ }\textbf {\bibinfo {volume} {04}},\ \bibinfo {pages} {078} (\bibinfo {year} {2016})},\ \Eprint {http://arxiv.org/abs/1511.09404} {arXiv:1511.09404 [hep-ph]} \BibitemShut {NoStop}%
	\bibitem [{\citenamefont {Gehrmann}\ \emph {et~al.}(2016)\citenamefont {Gehrmann}, \citenamefont {Henn},\ and\ \citenamefont {Lo~Presti}}]{Gehrmann:2015bfy}%
	  \BibitemOpen
	  \bibfield  {author} {\bibinfo {author} {\bibfnamefont {T.}~\bibnamefont {Gehrmann}}, \bibinfo {author} {\bibfnamefont {J.~M.}\ \bibnamefont {Henn}}, \ and\ \bibinfo {author} {\bibfnamefont {N.~A.}\ \bibnamefont {Lo~Presti}},\ }\href {\doibase 10.1103/PhysRevLett.116.062001} {\bibfield  {journal} {\bibinfo  {journal} {Phys. Rev. Lett.}\ }\textbf {\bibinfo {volume} {116}},\ \bibinfo {pages} {062001} (\bibinfo {year} {2016})},\ \bibinfo {note} {[Erratum: Phys.Rev.Lett. 116, 189903 (2016)]},\ \Eprint {http://arxiv.org/abs/1511.05409} {arXiv:1511.05409 [hep-ph]} \BibitemShut {NoStop}%
	\bibitem [{\citenamefont {Chicherin}\ \emph {et~al.}(2018{\natexlab{a}})\citenamefont {Chicherin}, \citenamefont {Henn},\ and\ \citenamefont {Mitev}}]{Chicherin:2017dob}%
	  \BibitemOpen
	  \bibfield  {author} {\bibinfo {author} {\bibfnamefont {D.}~\bibnamefont {Chicherin}}, \bibinfo {author} {\bibfnamefont {J.}~\bibnamefont {Henn}}, \ and\ \bibinfo {author} {\bibfnamefont {V.}~\bibnamefont {Mitev}},\ }\href {\doibase 10.1007/JHEP05(2018)164} {\bibfield  {journal} {\bibinfo  {journal} {JHEP}\ }\textbf {\bibinfo {volume} {05}},\ \bibinfo {pages} {164} (\bibinfo {year} {2018}{\natexlab{a}})},\ \Eprint {http://arxiv.org/abs/1712.09610} {arXiv:1712.09610 [hep-th]} \BibitemShut {NoStop}%
	\bibitem [{\citenamefont {Abreu}\ \emph {et~al.}(2019{\natexlab{a}})\citenamefont {Abreu}, \citenamefont {Dixon}, \citenamefont {Herrmann}, \citenamefont {Page},\ and\ \citenamefont {Zeng}}]{Abreu:2018aqd}%
	  \BibitemOpen
	  \bibfield  {author} {\bibinfo {author} {\bibfnamefont {S.}~\bibnamefont {Abreu}}, \bibinfo {author} {\bibfnamefont {L.~J.}\ \bibnamefont {Dixon}}, \bibinfo {author} {\bibfnamefont {E.}~\bibnamefont {Herrmann}}, \bibinfo {author} {\bibfnamefont {B.}~\bibnamefont {Page}}, \ and\ \bibinfo {author} {\bibfnamefont {M.}~\bibnamefont {Zeng}},\ }\href {\doibase 10.1103/PhysRevLett.122.121603} {\bibfield  {journal} {\bibinfo  {journal} {Phys. Rev. Lett.}\ }\textbf {\bibinfo {volume} {122}},\ \bibinfo {pages} {121603} (\bibinfo {year} {2019}{\natexlab{a}})},\ \Eprint {http://arxiv.org/abs/1812.08941} {arXiv:1812.08941 [hep-th]} \BibitemShut {NoStop}%
	\bibitem [{\citenamefont {Gehrmann}\ \emph {et~al.}(2018)\citenamefont {Gehrmann}, \citenamefont {Henn},\ and\ \citenamefont {Lo~Presti}}]{Gehrmann:2018yef}%
	  \BibitemOpen
	  \bibfield  {author} {\bibinfo {author} {\bibfnamefont {T.}~\bibnamefont {Gehrmann}}, \bibinfo {author} {\bibfnamefont {J.~M.}\ \bibnamefont {Henn}}, \ and\ \bibinfo {author} {\bibfnamefont {N.~A.}\ \bibnamefont {Lo~Presti}},\ }\href {\doibase 10.1007/JHEP10(2018)103} {\bibfield  {journal} {\bibinfo  {journal} {JHEP}\ }\textbf {\bibinfo {volume} {10}},\ \bibinfo {pages} {103} (\bibinfo {year} {2018})},\ \Eprint {http://arxiv.org/abs/1807.09812} {arXiv:1807.09812 [hep-ph]} \BibitemShut {NoStop}%
	\bibitem [{\citenamefont {Abreu}\ \emph {et~al.}(2019{\natexlab{b}})\citenamefont {Abreu}, \citenamefont {Page},\ and\ \citenamefont {Zeng}}]{Abreu:2018rcw}%
	  \BibitemOpen
	  \bibfield  {author} {\bibinfo {author} {\bibfnamefont {S.}~\bibnamefont {Abreu}}, \bibinfo {author} {\bibfnamefont {B.}~\bibnamefont {Page}}, \ and\ \bibinfo {author} {\bibfnamefont {M.}~\bibnamefont {Zeng}},\ }\href {\doibase 10.1007/JHEP01(2019)006} {\bibfield  {journal} {\bibinfo  {journal} {JHEP}\ }\textbf {\bibinfo {volume} {01}},\ \bibinfo {pages} {006} (\bibinfo {year} {2019}{\natexlab{b}})},\ \Eprint {http://arxiv.org/abs/1807.11522} {arXiv:1807.11522 [hep-th]} \BibitemShut {NoStop}%
	\bibitem [{\citenamefont {Chicherin}\ \emph {et~al.}(2019{\natexlab{a}})\citenamefont {Chicherin}, \citenamefont {Gehrmann}, \citenamefont {Henn}, \citenamefont {Lo~Presti}, \citenamefont {Mitev},\ and\ \citenamefont {Wasser}}]{Chicherin:2018mue}%
	  \BibitemOpen
	  \bibfield  {author} {\bibinfo {author} {\bibfnamefont {D.}~\bibnamefont {Chicherin}}, \bibinfo {author} {\bibfnamefont {T.}~\bibnamefont {Gehrmann}}, \bibinfo {author} {\bibfnamefont {J.~M.}\ \bibnamefont {Henn}}, \bibinfo {author} {\bibfnamefont {N.~A.}\ \bibnamefont {Lo~Presti}}, \bibinfo {author} {\bibfnamefont {V.}~\bibnamefont {Mitev}}, \ and\ \bibinfo {author} {\bibfnamefont {P.}~\bibnamefont {Wasser}},\ }\href {\doibase 10.1007/JHEP03(2019)042} {\bibfield  {journal} {\bibinfo  {journal} {JHEP}\ }\textbf {\bibinfo {volume} {03}},\ \bibinfo {pages} {042} (\bibinfo {year} {2019}{\natexlab{a}})},\ \Eprint {http://arxiv.org/abs/1809.06240} {arXiv:1809.06240 [hep-ph]} \BibitemShut {NoStop}%
	\bibitem [{\citenamefont {Chicherin}\ \emph {et~al.}(2019{\natexlab{b}})\citenamefont {Chicherin}, \citenamefont {Gehrmann}, \citenamefont {Henn}, \citenamefont {Wasser}, \citenamefont {Zhang},\ and\ \citenamefont {Zoia}}]{Chicherin:2018old}%
	  \BibitemOpen
	  \bibfield  {author} {\bibinfo {author} {\bibfnamefont {D.}~\bibnamefont {Chicherin}}, \bibinfo {author} {\bibfnamefont {T.}~\bibnamefont {Gehrmann}}, \bibinfo {author} {\bibfnamefont {J.~M.}\ \bibnamefont {Henn}}, \bibinfo {author} {\bibfnamefont {P.}~\bibnamefont {Wasser}}, \bibinfo {author} {\bibfnamefont {Y.}~\bibnamefont {Zhang}}, \ and\ \bibinfo {author} {\bibfnamefont {S.}~\bibnamefont {Zoia}},\ }\href {\doibase 10.1103/PhysRevLett.123.041603} {\bibfield  {journal} {\bibinfo  {journal} {Phys. Rev. Lett.}\ }\textbf {\bibinfo {volume} {123}},\ \bibinfo {pages} {041603} (\bibinfo {year} {2019}{\natexlab{b}})},\ \Eprint {http://arxiv.org/abs/1812.11160} {arXiv:1812.11160 [hep-ph]} \BibitemShut {NoStop}%
	\bibitem [{\citenamefont {Chicherin}\ and\ \citenamefont {Sotnikov}(2020)}]{Chicherin:2020oor}%
	  \BibitemOpen
	  \bibfield  {author} {\bibinfo {author} {\bibfnamefont {D.}~\bibnamefont {Chicherin}}\ and\ \bibinfo {author} {\bibfnamefont {V.}~\bibnamefont {Sotnikov}},\ }\href {\doibase 10.1007/JHEP12(2020)167} {\bibfield  {journal} {\bibinfo  {journal} {JHEP}\ }\textbf {\bibinfo {volume} {20}},\ \bibinfo {pages} {167} (\bibinfo {year} {2020})},\ \Eprint {http://arxiv.org/abs/2009.07803} {arXiv:2009.07803 [hep-ph]} \BibitemShut {NoStop}%
	\bibitem [{\citenamefont {Henn}(2013)}]{Henn:2013pwa}%
	  \BibitemOpen
	  \bibfield  {author} {\bibinfo {author} {\bibfnamefont {J.~M.}\ \bibnamefont {Henn}},\ }\href {\doibase 10.1103/PhysRevLett.110.251601} {\bibfield  {journal} {\bibinfo  {journal} {Phys. Rev. Lett.}\ }\textbf {\bibinfo {volume} {110}},\ \bibinfo {pages} {251601} (\bibinfo {year} {2013})},\ \Eprint {http://arxiv.org/abs/1304.1806} {arXiv:1304.1806 [hep-th]} \BibitemShut {NoStop}%
	\bibitem [{\citenamefont {Catani}(1998)}]{Catani:1998bh}%
	  \BibitemOpen
	  \bibfield  {author} {\bibinfo {author} {\bibfnamefont {S.}~\bibnamefont {Catani}},\ }\href {\doibase 10.1016/S0370-2693(98)00332-3} {\bibfield  {journal} {\bibinfo  {journal} {Phys. Lett.}\ }\textbf {\bibinfo {volume} {B427}},\ \bibinfo {pages} {161} (\bibinfo {year} {1998})},\ \Eprint {http://arxiv.org/abs/hep-ph/9802439} {arXiv:hep-ph/9802439 [hep-ph]} \BibitemShut {NoStop}%
	\bibitem [{\citenamefont {Sterman}\ and\ \citenamefont {Tejeda-Yeomans}(2003)}]{Sterman:2002qn}%
	  \BibitemOpen
	  \bibfield  {author} {\bibinfo {author} {\bibfnamefont {G.~F.}\ \bibnamefont {Sterman}}\ and\ \bibinfo {author} {\bibfnamefont {M.~E.}\ \bibnamefont {Tejeda-Yeomans}},\ }\href {\doibase 10.1016/S0370-2693(02)03100-3} {\bibfield  {journal} {\bibinfo  {journal} {Phys. Lett.}\ }\textbf {\bibinfo {volume} {B552}},\ \bibinfo {pages} {48} (\bibinfo {year} {2003})},\ \Eprint {http://arxiv.org/abs/hep-ph/0210130} {arXiv:hep-ph/0210130 [hep-ph]} \BibitemShut {NoStop}%
	\bibitem [{\citenamefont {Bern}\ \emph {et~al.}(1994)\citenamefont {Bern}, \citenamefont {Chalmers}, \citenamefont {Dixon},\ and\ \citenamefont {Kosower}}]{Bern:1993qk}%
	  \BibitemOpen
	  \bibfield  {author} {\bibinfo {author} {\bibfnamefont {Z.}~\bibnamefont {Bern}}, \bibinfo {author} {\bibfnamefont {G.}~\bibnamefont {Chalmers}}, \bibinfo {author} {\bibfnamefont {L.~J.}\ \bibnamefont {Dixon}}, \ and\ \bibinfo {author} {\bibfnamefont {D.~A.}\ \bibnamefont {Kosower}},\ }\href {\doibase 10.1103/PhysRevLett.72.2134} {\bibfield  {journal} {\bibinfo  {journal} {Phys. Rev. Lett.}\ }\textbf {\bibinfo {volume} {72}},\ \bibinfo {pages} {2134} (\bibinfo {year} {1994})},\ \Eprint {http://arxiv.org/abs/hep-ph/9312333} {arXiv:hep-ph/9312333} \BibitemShut {NoStop}%
	\bibitem [{\citenamefont {Kosower}(1999)}]{Kosower:1999xi}%
	  \BibitemOpen
	  \bibfield  {author} {\bibinfo {author} {\bibfnamefont {D.~A.}\ \bibnamefont {Kosower}},\ }\href {\doibase 10.1016/S0550-3213(99)00251-5} {\bibfield  {journal} {\bibinfo  {journal} {Nucl. Phys.}\ }\textbf {\bibinfo {volume} {B552}},\ \bibinfo {pages} {319} (\bibinfo {year} {1999})},\ \Eprint {http://arxiv.org/abs/hep-ph/9901201} {arXiv:hep-ph/9901201 [hep-ph]} \BibitemShut {NoStop}%
	\bibitem [{\citenamefont {Bern}\ \emph {et~al.}(2005)\citenamefont {Bern}, \citenamefont {Dixon},\ and\ \citenamefont {Smirnov}}]{Bern:2005iz}%
	  \BibitemOpen
	  \bibfield  {author} {\bibinfo {author} {\bibfnamefont {Z.}~\bibnamefont {Bern}}, \bibinfo {author} {\bibfnamefont {L.~J.}\ \bibnamefont {Dixon}}, \ and\ \bibinfo {author} {\bibfnamefont {V.~A.}\ \bibnamefont {Smirnov}},\ }\href {\doibase 10.1103/PhysRevD.72.085001} {\bibfield  {journal} {\bibinfo  {journal} {Phys. Rev.}\ }\textbf {\bibinfo {volume} {D72}},\ \bibinfo {pages} {085001} (\bibinfo {year} {2005})},\ \Eprint {http://arxiv.org/abs/hep-th/0505205} {arXiv:hep-th/0505205 [hep-th]} \BibitemShut {NoStop}%
	\bibitem [{\citenamefont {Anastasiou}\ \emph {et~al.}(2003)\citenamefont {Anastasiou}, \citenamefont {Bern}, \citenamefont {Dixon},\ and\ \citenamefont {Kosower}}]{Anastasiou:2003kj}%
	  \BibitemOpen
	  \bibfield  {author} {\bibinfo {author} {\bibfnamefont {C.}~\bibnamefont {Anastasiou}}, \bibinfo {author} {\bibfnamefont {Z.}~\bibnamefont {Bern}}, \bibinfo {author} {\bibfnamefont {L.~J.}\ \bibnamefont {Dixon}}, \ and\ \bibinfo {author} {\bibfnamefont {D.}~\bibnamefont {Kosower}},\ }\href {\doibase 10.1103/PhysRevLett.91.251602} {\bibfield  {journal} {\bibinfo  {journal} {Phys.Rev.Lett.}\ }\textbf {\bibinfo {volume} {91}},\ \bibinfo {pages} {251602} (\bibinfo {year} {2003})},\ \Eprint {http://arxiv.org/abs/hep-th/0309040} {arXiv:hep-th/0309040 [hep-th]} \BibitemShut {NoStop}%
	\bibitem [{\citenamefont {Goncharov}\ \emph {et~al.}(2010)\citenamefont {Goncharov}, \citenamefont {Spradlin}, \citenamefont {Vergu},\ and\ \citenamefont {Volovich}}]{Goncharov:2010jf}%
	  \BibitemOpen
	  \bibfield  {author} {\bibinfo {author} {\bibfnamefont {A.~B.}\ \bibnamefont {Goncharov}}, \bibinfo {author} {\bibfnamefont {M.}~\bibnamefont {Spradlin}}, \bibinfo {author} {\bibfnamefont {C.}~\bibnamefont {Vergu}}, \ and\ \bibinfo {author} {\bibfnamefont {A.}~\bibnamefont {Volovich}},\ }\href {\doibase 10.1103/PhysRevLett.105.151605} {\bibfield  {journal} {\bibinfo  {journal} {Phys. Rev. Lett.}\ }\textbf {\bibinfo {volume} {105}},\ \bibinfo {pages} {151605} (\bibinfo {year} {2010})},\ \Eprint {http://arxiv.org/abs/1006.5703} {arXiv:1006.5703 [hep-th]} \BibitemShut {NoStop}%
	\bibitem [{\citenamefont {Bern}\ \emph {et~al.}(2008)\citenamefont {Bern}, \citenamefont {Dixon}, \citenamefont {Kosower}, \citenamefont {Roiban}, \citenamefont {Spradlin}, \citenamefont {Vergu},\ and\ \citenamefont {Volovich}}]{Bern:2008ap}%
	  \BibitemOpen
	  \bibfield  {author} {\bibinfo {author} {\bibfnamefont {Z.}~\bibnamefont {Bern}}, \bibinfo {author} {\bibfnamefont {L.~J.}\ \bibnamefont {Dixon}}, \bibinfo {author} {\bibfnamefont {D.~A.}\ \bibnamefont {Kosower}}, \bibinfo {author} {\bibfnamefont {R.}~\bibnamefont {Roiban}}, \bibinfo {author} {\bibfnamefont {M.}~\bibnamefont {Spradlin}}, \bibinfo {author} {\bibfnamefont {C.}~\bibnamefont {Vergu}}, \ and\ \bibinfo {author} {\bibfnamefont {A.}~\bibnamefont {Volovich}},\ }\href {\doibase 10.1103/PhysRevD.78.045007} {\bibfield  {journal} {\bibinfo  {journal} {Phys. Rev. D}\ }\textbf {\bibinfo {volume} {78}},\ \bibinfo {pages} {045007} (\bibinfo {year} {2008})},\ \Eprint {http://arxiv.org/abs/0803.1465} {arXiv:0803.1465 [hep-th]} \BibitemShut {NoStop}%
	\bibitem [{\citenamefont {Drummond}\ \emph {et~al.}(2007)\citenamefont {Drummond}, \citenamefont {Henn}, \citenamefont {Smirnov},\ and\ \citenamefont {Sokatchev}}]{Drummond:2006rz}%
	  \BibitemOpen
	  \bibfield  {author} {\bibinfo {author} {\bibfnamefont {J.~M.}\ \bibnamefont {Drummond}}, \bibinfo {author} {\bibfnamefont {J.}~\bibnamefont {Henn}}, \bibinfo {author} {\bibfnamefont {V.~A.}\ \bibnamefont {Smirnov}}, \ and\ \bibinfo {author} {\bibfnamefont {E.}~\bibnamefont {Sokatchev}},\ }\href {\doibase 10.1088/1126-6708/2007/01/064} {\bibfield  {journal} {\bibinfo  {journal} {JHEP}\ }\textbf {\bibinfo {volume} {01}},\ \bibinfo {pages} {064} (\bibinfo {year} {2007})},\ \Eprint {http://arxiv.org/abs/hep-th/0607160} {arXiv:hep-th/0607160 [hep-th]} \BibitemShut {NoStop}%
	\bibitem [{\citenamefont {Bern}\ \emph {et~al.}(2018)\citenamefont {Bern}, \citenamefont {Enciso}, \citenamefont {Shen},\ and\ \citenamefont {Zeng}}]{Bern:2018oao}%
	  \BibitemOpen
	  \bibfield  {author} {\bibinfo {author} {\bibfnamefont {Z.}~\bibnamefont {Bern}}, \bibinfo {author} {\bibfnamefont {M.}~\bibnamefont {Enciso}}, \bibinfo {author} {\bibfnamefont {C.-H.}\ \bibnamefont {Shen}}, \ and\ \bibinfo {author} {\bibfnamefont {M.}~\bibnamefont {Zeng}},\ }\href {\doibase 10.1103/PhysRevLett.121.121603} {\bibfield  {journal} {\bibinfo  {journal} {Phys. Rev. Lett.}\ }\textbf {\bibinfo {volume} {121}},\ \bibinfo {pages} {121603} (\bibinfo {year} {2018})},\ \Eprint {http://arxiv.org/abs/arXiv:1806.06509} {arXiv:1806.06509} \BibitemShut {NoStop}%
	\bibitem [{\citenamefont {Chicherin}\ \emph {et~al.}(2018{\natexlab{b}})\citenamefont {Chicherin}, \citenamefont {Henn},\ and\ \citenamefont {Sokatchev}}]{Chicherin:2018wes}%
	  \BibitemOpen
	  \bibfield  {author} {\bibinfo {author} {\bibfnamefont {D.}~\bibnamefont {Chicherin}}, \bibinfo {author} {\bibfnamefont {J.~M.}\ \bibnamefont {Henn}}, \ and\ \bibinfo {author} {\bibfnamefont {E.}~\bibnamefont {Sokatchev}},\ }\href {\doibase 10.1007/JHEP09(2018)012} {\bibfield  {journal} {\bibinfo  {journal} {JHEP}\ }\textbf {\bibinfo {volume} {09}},\ \bibinfo {pages} {012} (\bibinfo {year} {2018}{\natexlab{b}})},\ \Eprint {http://arxiv.org/abs/1807.06321} {arXiv:1807.06321 [hep-th]} \BibitemShut {NoStop}%
	\bibitem [{\citenamefont {Korchemsky}\ and\ \citenamefont {Radyushkin}(1986)}]{Korchemsky:1985xj}%
	  \BibitemOpen
	  \bibfield  {author} {\bibinfo {author} {\bibfnamefont {G.~P.}\ \bibnamefont {Korchemsky}}\ and\ \bibinfo {author} {\bibfnamefont {A.~V.}\ \bibnamefont {Radyushkin}},\ }\bibfield  {booktitle} {\emph {\bibinfo {booktitle} {{International Seminar: Quarks 86 Tbilisi, USSR, April 15-17, 1986}}},\ }\href {\doibase 10.1016/0370-2693(86)91439-5} {\bibfield  {journal} {\bibinfo  {journal} {Phys. Lett.}\ }\textbf {\bibinfo {volume} {B171}},\ \bibinfo {pages} {459} (\bibinfo {year} {1986})}\BibitemShut {NoStop}%
	\bibitem [{\citenamefont {Korchemsky}(1989)}]{Korchemsky:1988si}%
	  \BibitemOpen
	  \bibfield  {author} {\bibinfo {author} {\bibfnamefont {G.~P.}\ \bibnamefont {Korchemsky}},\ }\href {\doibase 10.1142/S0217732389001453} {\bibfield  {journal} {\bibinfo  {journal} {Mod. Phys. Lett.}\ }\textbf {\bibinfo {volume} {A4}},\ \bibinfo {pages} {1257} (\bibinfo {year} {1989})}\BibitemShut {NoStop}%
	\bibitem [{\citenamefont {Beisert}\ \emph {et~al.}(2007)\citenamefont {Beisert}, \citenamefont {Eden},\ and\ \citenamefont {Staudacher}}]{Beisert:2006ez}%
	  \BibitemOpen
	  \bibfield  {author} {\bibinfo {author} {\bibfnamefont {N.}~\bibnamefont {Beisert}}, \bibinfo {author} {\bibfnamefont {B.}~\bibnamefont {Eden}}, \ and\ \bibinfo {author} {\bibfnamefont {M.}~\bibnamefont {Staudacher}},\ }\href {\doibase 10.1088/1742-5468/2007/01/P01021} {\bibfield  {journal} {\bibinfo  {journal} {J. Stat. Mech.}\ }\textbf {\bibinfo {volume} {0701}},\ \bibinfo {pages} {P01021} (\bibinfo {year} {2007})},\ \Eprint {http://arxiv.org/abs/hep-th/0610251} {arXiv:hep-th/0610251} \BibitemShut {NoStop}%
	\bibitem [{\citenamefont {Basso}(2012)}]{Basso:2010in}%
	  \BibitemOpen
	  \bibfield  {author} {\bibinfo {author} {\bibfnamefont {B.}~\bibnamefont {Basso}},\ }\href {\doibase 10.1016/j.nuclphysb.2011.12.010} {\bibfield  {journal} {\bibinfo  {journal} {Nucl. Phys. B}\ }\textbf {\bibinfo {volume} {857}},\ \bibinfo {pages} {254} (\bibinfo {year} {2012})},\ \Eprint {http://arxiv.org/abs/1010.5237} {arXiv:1010.5237 [hep-th]} \BibitemShut {NoStop}%
	\bibitem [{\citenamefont {Chicherin}\ \emph {et~al.}(2021)\citenamefont {Chicherin}, \citenamefont {Henn},\ and\ \citenamefont {Papathanasiou}}]{Chicherin:2020umh}%
	  \BibitemOpen
	  \bibfield  {author} {\bibinfo {author} {\bibfnamefont {D.}~\bibnamefont {Chicherin}}, \bibinfo {author} {\bibfnamefont {J.~M.}\ \bibnamefont {Henn}}, \ and\ \bibinfo {author} {\bibfnamefont {G.}~\bibnamefont {Papathanasiou}},\ }\href {\doibase 10.1103/PhysRevLett.126.091603} {\bibfield  {journal} {\bibinfo  {journal} {Phys. Rev. Lett.}\ }\textbf {\bibinfo {volume} {126}},\ \bibinfo {pages} {091603} (\bibinfo {year} {2021})},\ \Eprint {http://arxiv.org/abs/2012.12285} {arXiv:2012.12285 [hep-th]} \BibitemShut {NoStop}%
	\bibitem [{\citenamefont {Liu}(2022)}]{Liu:2022vck}%
	  \BibitemOpen
	  \bibfield  {author} {\bibinfo {author} {\bibfnamefont {Y.-T.}\ \bibnamefont {Liu}},\ }\href@noop {} {\  (\bibinfo {year} {2022})},\ \Eprint {http://arxiv.org/abs/2207.11815} {arXiv:2207.11815 [hep-th]} \BibitemShut {NoStop}%
	\bibitem [{\citenamefont {Caron-Huot}(2011{\natexlab{a}})}]{Caron-Huot:2011zgw}%
	  \BibitemOpen
	  \bibfield  {author} {\bibinfo {author} {\bibfnamefont {S.}~\bibnamefont {Caron-Huot}},\ }\href {\doibase 10.1007/JHEP12(2011)066} {\bibfield  {journal} {\bibinfo  {journal} {JHEP}\ }\textbf {\bibinfo {volume} {12}},\ \bibinfo {pages} {066} (\bibinfo {year} {2011}{\natexlab{a}})},\ \Eprint {http://arxiv.org/abs/1105.5606} {arXiv:1105.5606 [hep-th]} \BibitemShut {NoStop}%
	\bibitem [{\citenamefont {Golden}\ and\ \citenamefont {Spradlin}(2013)}]{Golden:2013lha}%
	  \BibitemOpen
	  \bibfield  {author} {\bibinfo {author} {\bibfnamefont {J.}~\bibnamefont {Golden}}\ and\ \bibinfo {author} {\bibfnamefont {M.}~\bibnamefont {Spradlin}},\ }\href {\doibase 10.1007/JHEP09(2013)111} {\bibfield  {journal} {\bibinfo  {journal} {JHEP}\ }\textbf {\bibinfo {volume} {09}},\ \bibinfo {pages} {111} (\bibinfo {year} {2013})},\ \Eprint {http://arxiv.org/abs/1306.1833} {arXiv:1306.1833 [hep-th]} \BibitemShut {NoStop}%
	\bibitem [{\citenamefont {Dixon}\ \emph {et~al.}(2023)\citenamefont {Dixon}, \citenamefont {G\"urdo\u{g}an}, \citenamefont {Liu}, \citenamefont {McLeod},\ and\ \citenamefont {Wilhelm}}]{Dixon:2022xqh}%
	  \BibitemOpen
	  \bibfield  {author} {\bibinfo {author} {\bibfnamefont {L.~J.}\ \bibnamefont {Dixon}}, \bibinfo {author} {\bibfnamefont {O.}~\bibnamefont {G\"urdo\u{g}an}}, \bibinfo {author} {\bibfnamefont {Y.-T.}\ \bibnamefont {Liu}}, \bibinfo {author} {\bibfnamefont {A.~J.}\ \bibnamefont {McLeod}}, \ and\ \bibinfo {author} {\bibfnamefont {M.}~\bibnamefont {Wilhelm}},\ }\href {\doibase 10.1103/PhysRevLett.130.111601} {\bibfield  {journal} {\bibinfo  {journal} {Phys. Rev. Lett.}\ }\textbf {\bibinfo {volume} {130}},\ \bibinfo {pages} {111601} (\bibinfo {year} {2023})},\ \Eprint {http://arxiv.org/abs/2212.02410} {arXiv:2212.02410 [hep-th]} \BibitemShut {NoStop}%
	\bibitem [{\citenamefont {Kotikov}\ and\ \citenamefont {Lipatov}(2003)}]{Kotikov:2002ab}%
	  \BibitemOpen
	  \bibfield  {author} {\bibinfo {author} {\bibfnamefont {A.~V.}\ \bibnamefont {Kotikov}}\ and\ \bibinfo {author} {\bibfnamefont {L.~N.}\ \bibnamefont {Lipatov}},\ }\href {\doibase 10.1016/S0550-3213(03)00264-5, 10.1016/j.nuclphysb.2004.02.032} {\bibfield  {journal} {\bibinfo  {journal} {Nucl. Phys.}\ }\textbf {\bibinfo {volume} {B661}},\ \bibinfo {pages} {19} (\bibinfo {year} {2003})},\ \bibinfo {note} {[Erratum: Nucl. Phys.B685,405(2004)]},\ \Eprint {http://arxiv.org/abs/hep-ph/0208220} {arXiv:hep-ph/0208220 [hep-ph]} \BibitemShut {NoStop}%
	\bibitem [{\citenamefont {Kotikov}\ \emph {et~al.}(2004)\citenamefont {Kotikov}, \citenamefont {Lipatov}, \citenamefont {Onishchenko},\ and\ \citenamefont {Velizhanin}}]{Kotikov:2004er}%
	  \BibitemOpen
	  \bibfield  {author} {\bibinfo {author} {\bibfnamefont {A.}~\bibnamefont {Kotikov}}, \bibinfo {author} {\bibfnamefont {L.}~\bibnamefont {Lipatov}}, \bibinfo {author} {\bibfnamefont {A.}~\bibnamefont {Onishchenko}}, \ and\ \bibinfo {author} {\bibfnamefont {V.}~\bibnamefont {Velizhanin}},\ }\href {\doibase 10.1016/j.physletb.2004.05.078, 10.1016/j.physletb.2004.05.078} {\bibfield  {journal} {\bibinfo  {journal} {Phys.Lett.}\ }\textbf {\bibinfo {volume} {B595}},\ \bibinfo {pages} {521} (\bibinfo {year} {2004})},\ \Eprint {http://arxiv.org/abs/hep-th/0404092} {arXiv:hep-th/0404092 [hep-th]} \BibitemShut {NoStop}%
	\bibitem [{\citenamefont {Jin}\ and\ \citenamefont {Yang}(2018)}]{Jin:2018fak}%
	  \BibitemOpen
	  \bibfield  {author} {\bibinfo {author} {\bibfnamefont {Q.}~\bibnamefont {Jin}}\ and\ \bibinfo {author} {\bibfnamefont {G.}~\bibnamefont {Yang}},\ }\href {\doibase 10.1103/PhysRevLett.121.101603} {\bibfield  {journal} {\bibinfo  {journal} {Phys. Rev. Lett.}\ }\textbf {\bibinfo {volume} {121}},\ \bibinfo {pages} {101603} (\bibinfo {year} {2018})},\ \Eprint {http://arxiv.org/abs/1804.04653} {arXiv:1804.04653 [hep-th]} \BibitemShut {NoStop}%
	\bibitem [{\citenamefont {Brandhuber}\ \emph {et~al.}(2017)\citenamefont {Brandhuber}, \citenamefont {Kostacinska}, \citenamefont {Penante},\ and\ \citenamefont {Travaglini}}]{Brandhuber:2017bkg}%
	  \BibitemOpen
	  \bibfield  {author} {\bibinfo {author} {\bibfnamefont {A.}~\bibnamefont {Brandhuber}}, \bibinfo {author} {\bibfnamefont {M.}~\bibnamefont {Kostacinska}}, \bibinfo {author} {\bibfnamefont {B.}~\bibnamefont {Penante}}, \ and\ \bibinfo {author} {\bibfnamefont {G.}~\bibnamefont {Travaglini}},\ }\href {\doibase 10.1103/PhysRevLett.119.161601} {\bibfield  {journal} {\bibinfo  {journal} {Phys. Rev. Lett.}\ }\textbf {\bibinfo {volume} {119}},\ \bibinfo {pages} {161601} (\bibinfo {year} {2017})},\ \Eprint {http://arxiv.org/abs/1707.09897} {arXiv:1707.09897 [hep-th]} \BibitemShut {NoStop}%
	\bibitem [{\citenamefont {Alday}\ \emph {et~al.}(2010)\citenamefont {Alday}, \citenamefont {Maldacena}, \citenamefont {Sever},\ and\ \citenamefont {Vieira}}]{Alday:2010vh}%
	  \BibitemOpen
	  \bibfield  {author} {\bibinfo {author} {\bibfnamefont {L.~F.}\ \bibnamefont {Alday}}, \bibinfo {author} {\bibfnamefont {J.}~\bibnamefont {Maldacena}}, \bibinfo {author} {\bibfnamefont {A.}~\bibnamefont {Sever}}, \ and\ \bibinfo {author} {\bibfnamefont {P.}~\bibnamefont {Vieira}},\ }\href {\doibase 10.1088/1751-8113/43/48/485401} {\bibfield  {journal} {\bibinfo  {journal} {J. Phys. A}\ }\textbf {\bibinfo {volume} {43}},\ \bibinfo {pages} {485401} (\bibinfo {year} {2010})},\ \Eprint {http://arxiv.org/abs/1002.2459} {arXiv:1002.2459 [hep-th]} \BibitemShut {NoStop}%
	\bibitem [{\citenamefont {Gao}\ and\ \citenamefont {Yang}(2013)}]{Gao:2013dza}%
	  \BibitemOpen
	  \bibfield  {author} {\bibinfo {author} {\bibfnamefont {Z.}~\bibnamefont {Gao}}\ and\ \bibinfo {author} {\bibfnamefont {G.}~\bibnamefont {Yang}},\ }\href {\doibase 10.1007/JHEP06(2013)105} {\bibfield  {journal} {\bibinfo  {journal} {JHEP}\ }\textbf {\bibinfo {volume} {06}},\ \bibinfo {pages} {105} (\bibinfo {year} {2013})},\ \Eprint {http://arxiv.org/abs/1303.2668} {arXiv:1303.2668 [hep-th]} \BibitemShut {NoStop}%
	\bibitem [{\citenamefont {Ouyang}\ and\ \citenamefont {Shu}(2022)}]{Ouyang:2022sje}%
	  \BibitemOpen
	  \bibfield  {author} {\bibinfo {author} {\bibfnamefont {H.}~\bibnamefont {Ouyang}}\ and\ \bibinfo {author} {\bibfnamefont {H.}~\bibnamefont {Shu}},\ }\href {\doibase 10.1007/JHEP05(2022)099} {\bibfield  {journal} {\bibinfo  {journal} {JHEP}\ }\textbf {\bibinfo {volume} {05}},\ \bibinfo {pages} {099} (\bibinfo {year} {2022})},\ \Eprint {http://arxiv.org/abs/2202.10700} {arXiv:2202.10700 [hep-th]} \BibitemShut {NoStop}%
	\bibitem [{\citenamefont {Mason}\ and\ \citenamefont {Skinner}(2010)}]{Mason:2010yk}%
	  \BibitemOpen
	  \bibfield  {author} {\bibinfo {author} {\bibfnamefont {L.~J.}\ \bibnamefont {Mason}}\ and\ \bibinfo {author} {\bibfnamefont {D.}~\bibnamefont {Skinner}},\ }\href {\doibase 10.1007/JHEP12(2010)018} {\bibfield  {journal} {\bibinfo  {journal} {JHEP}\ }\textbf {\bibinfo {volume} {12}},\ \bibinfo {pages} {018} (\bibinfo {year} {2010})},\ \Eprint {http://arxiv.org/abs/1009.2225} {arXiv:1009.2225 [hep-th]} \BibitemShut {NoStop}%
	\bibitem [{\citenamefont {Caron-Huot}(2011{\natexlab{b}})}]{Caron-Huot:2010ryg}%
	  \BibitemOpen
	  \bibfield  {author} {\bibinfo {author} {\bibfnamefont {S.}~\bibnamefont {Caron-Huot}},\ }\href {\doibase 10.1007/JHEP07(2011)058} {\bibfield  {journal} {\bibinfo  {journal} {JHEP}\ }\textbf {\bibinfo {volume} {07}},\ \bibinfo {pages} {058} (\bibinfo {year} {2011}{\natexlab{b}})},\ \Eprint {http://arxiv.org/abs/1010.1167} {arXiv:1010.1167 [hep-th]} \BibitemShut {NoStop}%
	\bibitem [{\citenamefont {Sever}\ \emph {et~al.}(2011)\citenamefont {Sever}, \citenamefont {Vieira},\ and\ \citenamefont {Wang}}]{Sever:2011da}%
	  \BibitemOpen
	  \bibfield  {author} {\bibinfo {author} {\bibfnamefont {A.}~\bibnamefont {Sever}}, \bibinfo {author} {\bibfnamefont {P.}~\bibnamefont {Vieira}}, \ and\ \bibinfo {author} {\bibfnamefont {T.}~\bibnamefont {Wang}},\ }\href {\doibase 10.1007/JHEP11(2011)051} {\bibfield  {journal} {\bibinfo  {journal} {JHEP}\ }\textbf {\bibinfo {volume} {11}},\ \bibinfo {pages} {051} (\bibinfo {year} {2011})},\ \Eprint {http://arxiv.org/abs/1108.1575} {arXiv:1108.1575 [hep-th]} \BibitemShut {NoStop}%
	\bibitem [{\citenamefont {Engelund}\ and\ \citenamefont {Roiban}(2013)}]{Engelund:2012re}%
	  \BibitemOpen
	  \bibfield  {author} {\bibinfo {author} {\bibfnamefont {O.~T.}\ \bibnamefont {Engelund}}\ and\ \bibinfo {author} {\bibfnamefont {R.}~\bibnamefont {Roiban}},\ }\href {\doibase 10.1007/JHEP03(2013)172} {\bibfield  {journal} {\bibinfo  {journal} {JHEP}\ }\textbf {\bibinfo {volume} {03}},\ \bibinfo {pages} {172} (\bibinfo {year} {2013})},\ \Eprint {http://arxiv.org/abs/1209.0227} {arXiv:1209.0227 [hep-th]} \BibitemShut {NoStop}%
	\bibitem [{\citenamefont {Brandhuber}\ \emph {et~al.}(2014)\citenamefont {Brandhuber}, \citenamefont {Penante}, \citenamefont {Travaglini},\ and\ \citenamefont {Wen}}]{Brandhuber:2014ica}%
	  \BibitemOpen
	  \bibfield  {author} {\bibinfo {author} {\bibfnamefont {A.}~\bibnamefont {Brandhuber}}, \bibinfo {author} {\bibfnamefont {B.}~\bibnamefont {Penante}}, \bibinfo {author} {\bibfnamefont {G.}~\bibnamefont {Travaglini}}, \ and\ \bibinfo {author} {\bibfnamefont {C.}~\bibnamefont {Wen}},\ }\href {\doibase 10.1007/JHEP08(2014)100} {\bibfield  {journal} {\bibinfo  {journal} {JHEP}\ }\textbf {\bibinfo {volume} {08}},\ \bibinfo {pages} {100} (\bibinfo {year} {2014})},\ \Eprint {http://arxiv.org/abs/1406.1443} {arXiv:1406.1443 [hep-th]} \BibitemShut {NoStop}%
	\bibitem [{\citenamefont {Wilhelm}(2015)}]{Wilhelm:2014qua}%
	  \BibitemOpen
	  \bibfield  {author} {\bibinfo {author} {\bibfnamefont {M.}~\bibnamefont {Wilhelm}},\ }\href {\doibase 10.1007/JHEP02(2015)149} {\bibfield  {journal} {\bibinfo  {journal} {JHEP}\ }\textbf {\bibinfo {volume} {02}},\ \bibinfo {pages} {149} (\bibinfo {year} {2015})},\ \Eprint {http://arxiv.org/abs/1410.6309} {arXiv:1410.6309 [hep-th]} \BibitemShut {NoStop}%
	\bibitem [{\citenamefont {Nandan}\ \emph {et~al.}(2015)\citenamefont {Nandan}, \citenamefont {Sieg}, \citenamefont {Wilhelm},\ and\ \citenamefont {Yang}}]{Nandan:2014oga}%
	  \BibitemOpen
	  \bibfield  {author} {\bibinfo {author} {\bibfnamefont {D.}~\bibnamefont {Nandan}}, \bibinfo {author} {\bibfnamefont {C.}~\bibnamefont {Sieg}}, \bibinfo {author} {\bibfnamefont {M.}~\bibnamefont {Wilhelm}}, \ and\ \bibinfo {author} {\bibfnamefont {G.}~\bibnamefont {Yang}},\ }\href {\doibase 10.1007/JHEP06(2015)156} {\bibfield  {journal} {\bibinfo  {journal} {JHEP}\ }\textbf {\bibinfo {volume} {06}},\ \bibinfo {pages} {156} (\bibinfo {year} {2015})},\ \Eprint {http://arxiv.org/abs/1410.8485} {arXiv:1410.8485 [hep-th]} \BibitemShut {NoStop}%
	\bibitem [{\citenamefont {Loebbert}\ \emph {et~al.}(2015)\citenamefont {Loebbert}, \citenamefont {Nandan}, \citenamefont {Sieg}, \citenamefont {Wilhelm},\ and\ \citenamefont {Yang}}]{Loebbert:2015ova}%
	  \BibitemOpen
	  \bibfield  {author} {\bibinfo {author} {\bibfnamefont {F.}~\bibnamefont {Loebbert}}, \bibinfo {author} {\bibfnamefont {D.}~\bibnamefont {Nandan}}, \bibinfo {author} {\bibfnamefont {C.}~\bibnamefont {Sieg}}, \bibinfo {author} {\bibfnamefont {M.}~\bibnamefont {Wilhelm}}, \ and\ \bibinfo {author} {\bibfnamefont {G.}~\bibnamefont {Yang}},\ }\href {\doibase 10.1007/JHEP10(2015)012} {\bibfield  {journal} {\bibinfo  {journal} {JHEP}\ }\textbf {\bibinfo {volume} {10}},\ \bibinfo {pages} {012} (\bibinfo {year} {2015})},\ \Eprint {http://arxiv.org/abs/1504.06323} {arXiv:1504.06323 [hep-th]} \BibitemShut {NoStop}%
	\bibitem [{\citenamefont {Brandhuber}\ \emph {et~al.}(2016)\citenamefont {Brandhuber}, \citenamefont {Kostacinska}, \citenamefont {Penante}, \citenamefont {Travaglini},\ and\ \citenamefont {Young}}]{Brandhuber:2016fni}%
	  \BibitemOpen
	  \bibfield  {author} {\bibinfo {author} {\bibfnamefont {A.}~\bibnamefont {Brandhuber}}, \bibinfo {author} {\bibfnamefont {M.}~\bibnamefont {Kostacinska}}, \bibinfo {author} {\bibfnamefont {B.}~\bibnamefont {Penante}}, \bibinfo {author} {\bibfnamefont {G.}~\bibnamefont {Travaglini}}, \ and\ \bibinfo {author} {\bibfnamefont {D.}~\bibnamefont {Young}},\ }\href {\doibase 10.1007/JHEP08(2016)134} {\bibfield  {journal} {\bibinfo  {journal} {JHEP}\ }\textbf {\bibinfo {volume} {08}},\ \bibinfo {pages} {134} (\bibinfo {year} {2016})},\ \Eprint {http://arxiv.org/abs/1606.08682} {arXiv:1606.08682 [hep-th]} \BibitemShut {NoStop}%
	\bibitem [{\citenamefont {Loebbert}\ \emph {et~al.}(2016)\citenamefont {Loebbert}, \citenamefont {Sieg}, \citenamefont {Wilhelm},\ and\ \citenamefont {Yang}}]{Loebbert:2016xkw}%
	  \BibitemOpen
	  \bibfield  {author} {\bibinfo {author} {\bibfnamefont {F.}~\bibnamefont {Loebbert}}, \bibinfo {author} {\bibfnamefont {C.}~\bibnamefont {Sieg}}, \bibinfo {author} {\bibfnamefont {M.}~\bibnamefont {Wilhelm}}, \ and\ \bibinfo {author} {\bibfnamefont {G.}~\bibnamefont {Yang}},\ }\href {\doibase 10.1007/JHEP12(2016)090} {\bibfield  {journal} {\bibinfo  {journal} {JHEP}\ }\textbf {\bibinfo {volume} {12}},\ \bibinfo {pages} {090} (\bibinfo {year} {2016})},\ \Eprint {http://arxiv.org/abs/1610.06567} {arXiv:1610.06567 [hep-th]} \BibitemShut {NoStop}%
	\bibitem [{\citenamefont {Banerjee}\ \emph {et~al.}(2017)\citenamefont {Banerjee}, \citenamefont {Dhani}, \citenamefont {Mahakhud}, \citenamefont {Ravindran},\ and\ \citenamefont {Seth}}]{Banerjee:2016kri}%
	  \BibitemOpen
	  \bibfield  {author} {\bibinfo {author} {\bibfnamefont {P.}~\bibnamefont {Banerjee}}, \bibinfo {author} {\bibfnamefont {P.~K.}\ \bibnamefont {Dhani}}, \bibinfo {author} {\bibfnamefont {M.}~\bibnamefont {Mahakhud}}, \bibinfo {author} {\bibfnamefont {V.}~\bibnamefont {Ravindran}}, \ and\ \bibinfo {author} {\bibfnamefont {S.}~\bibnamefont {Seth}},\ }\href {\doibase 10.1007/JHEP05(2017)085} {\bibfield  {journal} {\bibinfo  {journal} {JHEP}\ }\textbf {\bibinfo {volume} {05}},\ \bibinfo {pages} {085} (\bibinfo {year} {2017})},\ \Eprint {http://arxiv.org/abs/1612.00885} {arXiv:1612.00885 [hep-th]} \BibitemShut {NoStop}%
	\bibitem [{\citenamefont {Brandhuber}\ \emph {et~al.}(2018)\citenamefont {Brandhuber}, \citenamefont {Kostacinska}, \citenamefont {Penante},\ and\ \citenamefont {Travaglini}}]{Brandhuber:2018xzk}%
	  \BibitemOpen
	  \bibfield  {author} {\bibinfo {author} {\bibfnamefont {A.}~\bibnamefont {Brandhuber}}, \bibinfo {author} {\bibfnamefont {M.}~\bibnamefont {Kostacinska}}, \bibinfo {author} {\bibfnamefont {B.}~\bibnamefont {Penante}}, \ and\ \bibinfo {author} {\bibfnamefont {G.}~\bibnamefont {Travaglini}},\ }\href {\doibase 10.1007/JHEP12(2018)076} {\bibfield  {journal} {\bibinfo  {journal} {JHEP}\ }\textbf {\bibinfo {volume} {12}},\ \bibinfo {pages} {076} (\bibinfo {year} {2018})},\ \Eprint {http://arxiv.org/abs/1804.05703} {arXiv:1804.05703 [hep-th]} \BibitemShut {NoStop}%
	\bibitem [{\citenamefont {Lin}\ and\ \citenamefont {Yang}(2021)}]{Lin:2020dyj}%
	  \BibitemOpen
	  \bibfield  {author} {\bibinfo {author} {\bibfnamefont {G.}~\bibnamefont {Lin}}\ and\ \bibinfo {author} {\bibfnamefont {G.}~\bibnamefont {Yang}},\ }\href {\doibase 10.1007/JHEP04(2021)176} {\bibfield  {journal} {\bibinfo  {journal} {JHEP}\ }\textbf {\bibinfo {volume} {04}},\ \bibinfo {pages} {176} (\bibinfo {year} {2021})},\ \Eprint {http://arxiv.org/abs/2011.06540} {arXiv:2011.06540 [hep-th]} \BibitemShut {NoStop}%
	\bibitem [{\citenamefont {Basso}\ and\ \citenamefont {Tumanov}(2023)}]{Basso:2023bwv}%
	  \BibitemOpen
	  \bibfield  {author} {\bibinfo {author} {\bibfnamefont {B.}~\bibnamefont {Basso}}\ and\ \bibinfo {author} {\bibfnamefont {A.~G.}\ \bibnamefont {Tumanov}},\ }\href@noop {} {\  (\bibinfo {year} {2023})},\ \Eprint {http://arxiv.org/abs/2308.08432} {arXiv:2308.08432 [hep-th]} \BibitemShut {NoStop}%
	\bibitem [{\citenamefont {Hidding}(2021)}]{Hidding:2020ytt}%
	  \BibitemOpen
	  \bibfield  {author} {\bibinfo {author} {\bibfnamefont {M.}~\bibnamefont {Hidding}},\ }\href {\doibase 10.1016/j.cpc.2021.108125} {\bibfield  {journal} {\bibinfo  {journal} {Comput. Phys. Commun.}\ }\textbf {\bibinfo {volume} {269}},\ \bibinfo {pages} {108125} (\bibinfo {year} {2021})},\ \Eprint {http://arxiv.org/abs/2006.05510} {arXiv:2006.05510 [hep-ph]} \BibitemShut {NoStop}%
	\bibitem [{\citenamefont {Liu}\ and\ \citenamefont {Ma}(2022)}]{Liu:2022chg}%
	  \BibitemOpen
	  \bibfield  {author} {\bibinfo {author} {\bibfnamefont {X.}~\bibnamefont {Liu}}\ and\ \bibinfo {author} {\bibfnamefont {Y.-Q.}\ \bibnamefont {Ma}},\ }\href@noop {} {\  (\bibinfo {year} {2022})},\ \Eprint {http://arxiv.org/abs/2201.11669} {arXiv:2201.11669 [hep-ph]} \BibitemShut {NoStop}%
	\bibitem [{\citenamefont {Gaiotto}\ \emph {et~al.}(2011)\citenamefont {Gaiotto}, \citenamefont {Maldacena}, \citenamefont {Sever},\ and\ \citenamefont {Vieira}}]{Gaiotto:2011dt}%
	  \BibitemOpen
	  \bibfield  {author} {\bibinfo {author} {\bibfnamefont {D.}~\bibnamefont {Gaiotto}}, \bibinfo {author} {\bibfnamefont {J.}~\bibnamefont {Maldacena}}, \bibinfo {author} {\bibfnamefont {A.}~\bibnamefont {Sever}}, \ and\ \bibinfo {author} {\bibfnamefont {P.}~\bibnamefont {Vieira}},\ }\href {\doibase 10.1007/JHEP12(2011)011} {\bibfield  {journal} {\bibinfo  {journal} {JHEP}\ }\textbf {\bibinfo {volume} {12}},\ \bibinfo {pages} {011} (\bibinfo {year} {2011})},\ \Eprint {http://arxiv.org/abs/1102.0062} {arXiv:1102.0062 [hep-th]} \BibitemShut {NoStop}%
\end{thebibliography}
\end{document}